\documentclass[fleqn,usenatbib]{mnras}

\usepackage{newtxtext,newtxmath}

\usepackage[T1]{fontenc}

\DeclareRobustCommand{\VAN}[3]{#2}
\let\VANthebibliography\thebibliography
\def\thebibliography{\DeclareRobustCommand{\VAN}[3]{##3}\VANthebibliography}

\usepackage{graphicx}	
\usepackage{amsmath}	
\usepackage{color,soul}

\numberwithin{equation}{section}

\title[Density \& metallicity of the MW's gas halo]{Constraining density and metallicity of the Milky Way's hot gas halo from \ion{O}{vii} spectra and ram-pressure stripping}

\author[N. Martynenko]{
Nickolay Martynenko$^{1,2}$\thanks{E-mail: martynenko.ns18@physics.msu.ru, martynenko@ms2.inr.ac.ru}
\\
$^{1}$Faculty of Physics, Lomonosov Moscow State University, 1-2 Leninskiye Gory, 119991, Moscow, Russia\\
$^{2}$Institute for Nuclear Research of the Russian Academy of Sciences, 60th October Anniversary Prospect 7a, Moscow 117312, Russia
}

\date{Accepted 2022 January 15. Received 2022 January 2; in original form 2021 May 11}

\pubyear{2022}

\begin{document}
\label{firstpage}
\pagerange{\pageref{firstpage}--\pageref{lastpage}}
\maketitle

\begin{abstract}
Milky Way's hot gaseous halo extends up to the Galactic virial radius ($\sim 200$ kpc) and contains a significant component of baryon mass of the Galaxy. The halo properties can be constrained from X-ray spectroscopic observations and from satellite galaxies' ram-pressure stripping studies. Results of the former method crucially depend on the gas metallicity assumptions while the latter one's are insensitive to them. Here, a joint analysis of both kinds of data is presented to constrain electron density and metallicity of the gas. The power law is assumed for the electron density radial profile, while for the metallicity, a common-used constant-metallicity assumption is relaxed by introducing of a physically motivated spherical profile. The model is fitted to a sample of 431 (18) sight lines for \ion{O}{vii} emission (absorption) measurements and 7 electron density constraints from ram-pressure stripping studies. The best-fitting halo-associated electron density profile of $n_e\propto r^{-(0.9...1.1)}$ (where $r\gg1$ kpc is the galactocentric radius) is found. The metallicity is constrained as $Z\simeq(0.1...0.7)~Z_{\sun}$ (subscript $\sun$ represents the solar values) at $r\ga50$ kpc. These imply a total hot gas mass of $M\simeq (2.4...8.7) \times10^{10}~\text{M}_{\sun}$, which accounts for $\sim(17...100)$ per cent of the Milky Way's missing baryon mass. The model uncertainties are discussed, and the results are examined in the context of previous studies.
\end{abstract}




\section{Introduction}

The Milky Way (MW) circumgalactic medium (CGM) extends up to the Galactic virial radius ($R_{\text{vir}}\sim200$~kpc) and is filled with diffuse gaseous matter. Due to its large volume, CGM is considered to contain a significant component of the Galactic mass, and this has been confirmed by a number of recent studies \citep[e. g.][]{AndersonBregman2010,MillerBregman2013,MillerBregman2015,Nuza2014,Troitsky2017,Kaaretetal2020}. It makes CGM a subject of interest in terms of the missing baryon problem, a local Universe deficit of observed baryon matter in comparison with the amount implied by cosmology \citep[see e.~g.][]{AndersonBregman2010,Guptaetal2012,Troitsky2017,BregmanAnderson2018}.

In addition, CGM is expected to have a significant contribution to the diffuse gamma-ray and neutrino backgrounds  \citep[][]{Feldmann2013,Taylor2014,KalashevTroitsky2016,Liuetal2019,Gabici2021} because of interactions of the circumgalactic gas with cosmic rays. 

There are several methods to constrain CGM gas density profile. On the one hand, the ram pressure of the gas strips dwarf galaxies moving inside CGM of their own gas. By simulations of this process together with satellite galaxies observations, the halo density can be estimated without any assumptions on the gas chemical composition \citep[see][]{BlitzRobishaw2000,GrcevichPutman2009,Gattoetal2013,Salemetal2015,Putmanetal2021}.

On the other hand, the density can be constrained from X-ray observations detecting a zero-redshift gas in absorption spectra of distant background sources \citep[][]{Yaoetal2012,Guptaetal2012,MillerBregman2013,Fang2015} and in blank-sky emission spectra \citep[][]{SnowdenEgger1997,McCammon2002,HenleyShelton2012,MillerBregman2015,Kaaretetal2020}. This approach requires additional assumptions, in particular, on the yet unknown gas composition usually encoded in its metallicity~$Z$. The research to date has however tended to focus on the density rather than metallicity. In most of the recent studies the latter parameter is considered constant, but the value is estimated very roughly, which implies results of large uncertainty and limited applicability. Moreover, sometimes the assumptions are somewhat contradictory. E. g., in the work of \citet{MillerBregman2015} the authors describe the gas emission properties referring to the work by \citet{Smith2001}, where the solar abundances of elements were assumed. None the less, they then estimate the metallicity as $Z \simeq 0.3~Z_{\sun}$ from the results. This implies the electron density to have been underestimated, since in the considered model $n_e \propto Z^{-1}$. This, however, generates unclear uncertainty in both properties and thus the results could have been distorted.  

It is also important to note that the density profiles derived by ram-pressure stripping observations and X-ray spectroscopic studies separately are not consistent. In the work by~\citet{Troitsky2017}, it has been demonstrated that the discrepancy between data can be resolved by introducing a metallicity profile. It was speculatively assumed that the metallicity profile is similar to that of electron density which entails a somewhat mimicry of the previously derived density profile by the true profiles of density and metallicity.

In the work of~\citet{Voit2019}, a metallicity gradient of $Z \propto r^{-0.5}$ inspired by observations of the relation between X-ray luminosity and gas temperature in galaxy groups was assumed in order to model column densities of highly ionized oxygen associated with the Galactic halo. However, a systematic understanding of how to describe and physically motivate the metallicity profile is still lacking.

The present study analyses observations of ram-pressure stripping and \ion{O}{vii} spectra associated with CGM jointly, aiming to provide a physical, self-consistent description of these data. The halo parametric model developed in \citet{MillerBregman2013, MillerBregman2015}, as well as the spectroscopic data filtering procedure presented by \citet{MillerBregman2015}, is modified and improved in this work, primarily, by relaxing a constant-metallicity assumption and introducing a physically motivated metallicity profile. 

In contrast to earlier studies and following to the approach proposed by~\citet{Troitsky2017}, the model is fitted to the pre-filtered sample of both spectroscopic and ram-pressure stripping observations.

It is not the task of this paper to examine whether or not the introduced metallicity profile is optimal, and whether such an improvement is necessary to achieve self-consistency. It is hoped, however, that this research will contribute to a deeper understanding of how the metallicity assumptions affect the inferred CGM physical properties and provide new insights into how the halo gas chemical composition can be constrained by the observations. 

The rest of the paper is organized as follows. Section~\ref{sec:data} describes the data selection and filtering. In Subsection~\ref{subsec:supplementarydata}, supplementary observational constraints used in the fitting procedure are discussed. In Section~\ref{sec:methodsresults}, the parametric model (Subsection~\ref{subsec:modeldescription}) is developed, the fitting procedure (Subsection~\ref{subsec:modelfitting}) is described, and the results (Subsection~\ref{subsec:results}) are briefly presented. Section~\ref{sec:discussion} compares the results with previous studies (Subsection~\ref{subsec:electrondensity_metallicity}) and discusses their limitations and stability (Subsection~\ref{subsec:limitations}), self-consistency (Subsection~\ref{subsec:selfconsistency}), and implications for the physical properties of the MW (Subsection~\ref{subsec:implications}). Finally, Section~\ref{sec:conclusions} provides a summary of this work.

\section{Data}\label{sec:data}
This section outlines measurement methods, discusses the relationship between the measured physical quantities and that of interest for this study, describes the data reduction procedure and presents supplementary observational constraints. Since the purpose of this investigation is not to review but to self-consistently analyse data, a detailed description of how the samples were initially obtained is not provided here, and the reader is referred to the original works. Instead, the more important issues in terms of data analysis are paid attention to. Specifically, possible sample contamination, dominant sources of uncertainty and limitations of the constraints are discussed. 

The spectroscopic observations used here are from \textit{XMM-Newton} archival data previously listed in the works on the CGM gas, and the ram-pressure stripping constraints are inferred from \ion{H}{i} detection together with hydrodynamic simulations.

\subsection{Ram-pressure stripping}\label{subsec:rampressure}
Data from several Galactic \ion{H}{i} surveys have identified that \ion{H}{i} content of the Local Group dwarf galaxies exhibits a decline if the galaxy is within $\sim (250...270)$ kpc of the MW or M31 \citep[see e. g.][]{BlitzRobishaw2000}. A number of studies have attempted to explain these observations in terms of ram-pressure stripping, the process of gas loss due to the pressure exerted by the gas of the host galaxy halo. This process can be briefly described as follows.

The condition under which the ambient pressure causes a disc dwarf galaxy with a relative speed of $v$ to lose gas can be written as:
\begin{equation}\label{eq:rampressure}
    \rho v^2 \ga 2 \pi G (\Sigma_* (\mathcal{R}) + \varsigma \Sigma_g (\mathcal{R}) ) \Sigma_g (\mathcal{R})
\end{equation}
In this equation, $G$ is the gravitational constant, $\rho$ is the halo mass density, $\Sigma_*$ and $\Sigma_g$ are, respectively, surface densities of dwarf's stellar matter and gas at the given radius (i. e. distance to the centre of disc) $\mathcal{R}$,  and $0\leq\varsigma\leq1$ is a constant representing a degree of dwarf's gas gravitational self-interaction.

As can be seen from~(\ref{eq:rampressure}), the higher halo density implies the more intensive stripping. Therefore, since the halo density is expected to decrease with galactocentric distance, it is reasonable to suppose the most sufficient stripping to occur at the pericentre of a satellite. This allows to estimate the mass density of the CGM at the pericentre by assuming ram-pressure stripping as a dominant contributor to the gas loss, calculating the orbit parameters, and estimating the initial gas amount and distribution in the satellite.

These considerations apply not only to disc satellites but also to spheroidal and irregular dwarfs. Several recent studies have modelled ram-pressure stripping in the MW CGM using observations and thus constrained the halo density at different galactocentric radii \citep[][]{GrcevichPutman2009,Gattoetal2013,Salemetal2015}.

The dominant sources of uncertainty in this approach are the kinematic parameters of the satellite, which determine its velocity and the orbit pericentre and thus crucially affect the resulting halo density constraint. \citet{Troitsky2017} has noted some errors to be on the order of the measured values and excluded these from the model fitting. 

This work uses the same sample of constraints as that listed by~\citet{Troitsky2017}, see Table~\ref{tab:RamPressure}. However, the fitting procedure does not exclude any of them since these data directly constrain the halo density profile without any assumptions on the gas metallicity, which is remarkably important for the quality of this analysis.

It is also notable that, to the best of the author's knowledge, there are no more recent simulations on the ram-pressure stripping for the MW dwarf satellites to date. The orbits' parameters used in the simulations referred to here are likely to be biased estimates, even inconsistent with the recent observational constraints \citep[see e.~g.][]{Hefan2021}, which implies a systematic uncertainty in both pericentres and electron density estimates not considered by this work.

Considering the parameters reported by~\citet{Hefan2021}, there is $\simeq (3...4)\sigma$ discrepancy for the Carina dwarf pericentres, and the other pericentres are consistent within $1\sigma$ with their improved values. However, in order to account for this properly, re-simulations are needed.

In this paper, ram-pressure stripping is assumed as the only mechanism of the gas loss, which might be a rough approximation. There is a consensus that tidal effects are likely to be unimportant \citep[see e.~g.][]{BlitzRobishaw2000,Gattoetal2013,Salemetal2015}, but a comprehensive quantitative analysis of all the possible mechanisms is still lacking, and is deferred here for further studies.
\begin{table}
	\centering
	\caption{Constraints on the gas electron density $n_e$ at the perigalacticons of the dwarf satellites. In brackets, 68 per cent confidence level (CL) intervals are shown. References: (a) \citet{GrcevichPutman2009}, (b) \citet{Gattoetal2013}, (c) \citet{Salemetal2015}}
	\label{tab:RamPressure}
    \begin{tabular}{l c c c}
        \hline 
        \hline
         Dwarf satellite & $r_{\text{peri}}$, kpc & $n_e$, 10$^{-4}$ cm$^{-3}$ & Reference\\
         \hline
         Carina & 20 (9.7...46.1) & 0.85 (0.67...2.70) & (a)\\
         Ursa Minor & 40 (21.8...61.9) & 2.10 (0.90...5.20)& \\
         Sculptor & 68 (45.5...77.1) & 2.70 (1.37...3.43)& \\
         Fornax & 118 (86.4...133.8) & 3.10 (1.81...4.01)& \\
         \hline
         Sextans & 73.5 (59.8...90.2) & 0.86 (0.62...2.38) & (b)\\
         Carina & 64.7 (51.2...81.8) & 0.81 (0.71...1.71) & \\
         \hline
         LMC & 48.2 (43.2...53.2) & 1.10 (0.65...1.54) & (c)\\
         \hline
        \end{tabular}
\end{table}

\subsection{Oxygen spectra}\label{subsec:Oviispectra}
Numerous X-ray spectroscopic observations have detected emitting \citep[][]{SnowdenEgger1997,McCammon2002,HenleyShelton2012,MillerBregman2015,Kaaretetal2020} and absorbing \citep[][]{Yaoetal2012,Guptaetal2012,MillerBregman2013,Fang2015} medium at zero-redshift, which is likely to be associated with a hot gaseous halo of the MW. Since each individual sight line probes the emitting/absorbing gas properties along the given direction, a large sample of spectroscopic measurements allows to constrain global properties of the gas, including the density profile. 

A major disadvantage of this method is that the results are very sensitive to assumptions about the chemical composition of the gas. Spectroscopic observations can only probe the number of the emitting or absorbing ions along the given line of sight. Therefore, to derive the density profile, the fraction of these ions in the gas must be known or assumed. Usually, the composition of the gas is encoded in its metallicity $Z$, the mass-fraction of elements that are heavier than hydrogen and helium. In this work, this chemical composition representation is also used, assuming that possible deviations in the relative abundances affect the results negligibly.  

Sample contamination is another substantial source of uncertainty. Beside the signal producer of spectral lines, the halo gas, there are also various noise sources. These include both local (such as solar radiation and near-solar interstellar gas) and distant (such as high-velocity clouds, HVCs) sources. Although all these objects are the parts of Galactic substructure, the current study distinguishes them from the halo and thus aims to reduce their contributions to the spectroscopic data. The filtering procedure is intended to extract the data constraining only the background halo-associated density profile, excluding all the contribution of non-spherical features (most importantly, the region near the MW disc).

\subsubsection{Absorption lines}
The circumgalactic gas reveals itself in absorption spectra of distant background sources, specifically, active galactic nuclei (AGNs). The absorption line at $z\simeq0$ detected in the spectrum of a source with a non-zero redshift is reasonable to be associated with the MW halo gas \citep[see, however,][where the possibility of cancelling of the redshift effect by the AGN outflow velocity is discussed]{Fang2015}. Since the absorption line equivalent width depends on the number of absorbing ions between the source and the observer, measuring AGNs' spectra allows to constrain the gas density profile.

This study uses data from the works by \citet{MillerBregman2013, Fang2015} (see Table~1, columns~10--11 and Table~1, column~9 in the articles, respectively). The authors analysed \textit{XMM-Newton} archival observations of AGNs' absorption spectra and derived \ion{O}{vii} column densities (the number of ions per unit area along the given direction) for 26 and 43 individual sight lines, respectively. In the work by \citet{MillerBregman2013}, two Galactic sources and one source in the Large Magellanic Cloud (LMC) were also considered. However, these are not included in this analysis with the intention to study the global halo properties exclusively, with no respect to observations probing the individual inner regions of the CGM.

Some of the sight lines are presented in the both works. In this case, the data from the work by \citet{Fang2015} are preferred for the current analysis, since the authors fitted the Doppler width as a free parameter for each individual sight line. \citet{MillerBregman2013}, in contrast, fixed the Doppler width at~150~km~s$^{-1}$. Therefore, the former procedure is expected to have given more accurate results.

\subsubsection{Emission lines}
The circumgalactic gas can be constrained by the blank-sky emission spectrum, i. e. diffuse X-ray background that is not due to resolved discrete sources. Methodically, this approach is less accurate than that based on absorption spectra. The observed diffuse emission contains contributions from all possible sources that are challenging to be distinguished. Hence, to conduct an acceptable analysis, it is necessary to properly estimate the halo and non-halo contributions to the line intensities and the corresponding degree of uncertainty.

In the current analysis, the all-sky \ion{O}{vii} emission data catalogued in~\citet{HenleyShelton2012} and previously referred to in the analysis by \citet{MillerBregman2015} are used (see also the latter work for a brief description of the sample and data filtering). Based on the \textit{XMM-Newton} archival data, the catalog presents line strengths for $\sim1000$ individual sight lines from the entire sky. In the original article, the intensities are measured in photons~cm$^{-2}$~s$^{-1}$~sr$^{-1}$, which is hereafter referred to as line units (L.U.).

\citet{HenleyShelton2012} filtered the sample by reducing the contamination from solar wind charge exchange emission. The article presents a sample with the data which is likely to have been contaminated excluded, as well as the non-filtered sample (see Table~2 and Table~1, respectively). The current study only uses the filtered sample.

\begin{figure}
	\includegraphics[width=\columnwidth]{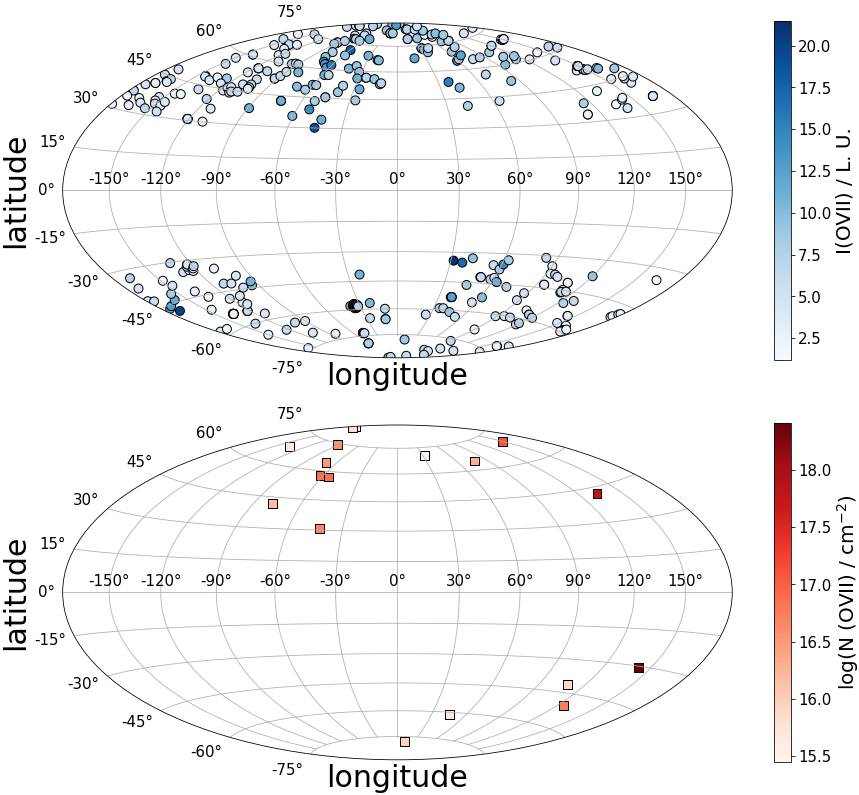}
    \caption{Filtered sample of \ion{O}{vii} emission (top) and absorption (bottom) lines. Latitudes and longitudes are Galactic. See Sub-subsection~\ref{subsubsec:filtering} for a detailed discussion.}
    \label{fig:oxygenspectra}
\end{figure}

\subsubsection{Data filtering}\label{subsubsec:filtering}
This work aims to constrain the radial density profile of the MW's extended gas halo. Therefore, the data contribution that is due to asymmetry of the halo substructure, including the Galactic disc, should be reduced. 

For self-consistency of the analysis, the absorption and emission line samples are filtered by the same criteria, even though some of these are only necessary for one type of observations and too strict for the other. 

To reduce the contamination from the Galactic disc, the sight lines with the galactic latitudes $|b|<30\degr$ are excluded. The sight line path through the MW disc matter is thus limited to the disc characteristic thickness ($\simeq0.3$ kpc), which is much less than the radius of the halo. The possible contamination of the sample also might be associated with the central region of the Galaxy, in particular, Fermi Bubbles and the MW bulge. These objects' structure examination is beyond the scope of the current analysis, and the sight lines within the galactic longitudes of $0\degr<l<10\degr$ and $350\degr<l<360\degr$ are also excluded.  

After excluding the regions discussed above, the sample is filtered more selectively. For this purpose, the~\ion{H}{i} column density map presented in the work by~\citet{Westmeier2018} and based on the all-sky \textit{HI4PI} survey (see Figure A1 in the paper) is used. By visual inspection, sight lines that pass through the high-density regions ($\log \left(N_{\text{\ion{H}{i}}}/\text{cm}^{-2}\right) \ga 19$), including areas near the Large and Small Magellanic Clouds, are excluded individually. Analogouysly, using the map presented by~\citet{Ackermannetal2014} (see Figure~5 in the paper), the region which is possibly contaminated by the Fermi Bubbles emission is also excluded.

The number of observations in the absorption (emission) lines sample is thereby reduced to 18 (431) (see Figure~\ref{fig:oxygenspectra}).

It is important to emphasize that the discussed filtering procedure is significantly limiting for the current analysis, since a vast region of the halo corresponding to the sight lines with above-average density of the gas is intentionally excluded from the consideration. Therefore, the further result should be interpreted with caution. The density profile to be obtained would rather characterise the background density of the halo and the radial distribution of the gas with no respect to the substructure of the CGM and underestimate the true amount of the hot gas. In particular, the density must be significantly underestimated near the MW disc, where the true amount of gas is determined by the sum of the background spherical distribution associated with the halo and the dominant axial symmetric distribution associated with the disc.

Also, there is a potential for bias from the fact that the reduced sample angular distribution is not isotropic (the median values of $|b|$ and $l$ are $55.1\degr$ and $210.6\degr$, respectively), and the sample of emission lines is much larger than that of absorption lines. 

\subsection{Supplementary constraints}\label{subsec:supplementarydata}
In order to improve the analysis quality and to make it easier to find the optimal parameters of the model (see Subsection \ref{subsec:modelfitting}), additional observational constraints on the density of the gas are used.

In an analysis of ram-pressure stripping, \citet{BlitzRobishaw2000} derived a lower bound on the halo density averaged over a volume of 250 kpc cutoff radius. Referring to the studies dealing with \ion{H}{i} detection in the dwarf galaxies within the cutoff radius of the MW, the authors determined the largest of the satellite-centred radii to which~\ion{H}{i} is observed. Considering the ram-pressure stripping as a dominant gas-loss mechanism, they concluded the halo number density to be $\ga2.5 \times 10^{-5}$~cm$^{-3}$. Thus, the electron density averaged over a volume of the Galactic virial radius (which is supposed to be less than 250 kpc, see Subsection \ref{subsec:modeldescription}) must be $\ga1.3\times10^{-5}$~cm$^{-3}$.

In a study by \citet{AndersonBregman2010}, an upper bound on the halo density was derived. The constraint comes from the dispersion measure (DM) of pulsars in the LMC. The authors analysed the DM of 11 pulsars in the direction of LMC presented in the work by \citet{Manchester2006} and concluded three of these sources to be located inside the MW. Thus, they estimated Galactic contribution to the DM, and subtracted it from the total DM of the pulsars that were associated with the LMC. Assuming a distance to the LMC of 50 kpc from the Sun, the sight-line-averaged halo electron density in this direction is constrained as $\la5 \times 10^{-4}$ cm$^{-3}$.

\section{Methods \& results}\label{sec:methodsresults}
This section describes the parametric model and the fitting procedure, and briefly presents the results. This study rather improves the model previously developed in the works by \citet{MillerBregman2013, MillerBregman2015} than develops a completely new one. Thus, the section primarily focuses on the improvements and provides only a brief discussion for the unchanged elements.

\subsection{Model description}\label{subsec:modeldescription}
The MW halo is assumed to be spherical with a cutoff radius of $R_{\text{vir}}=223$~kpc. The radius corresponds to the median of 11 recent estimates of the MW virial mass based on the data from \textit{Gaia} DR2 and presented in the work by~\citet{Wang2020}.  

Physical properties of the halo are encoded in two radial functions, the electron density profile and the metallicity profile. All the substructure of the halo is ignored, except for the Local Bubble (LB). The latter is taken into account when modelling the oxygen spectrum.

\subsubsection{Electron density}
To be consistent with previous studies, a spherical $\beta$-model is assumed for the electron density. The use of this model is initially motivated by the fact that it well reproduces the observed X-ray surface brightness profile of other galaxies \citep[see e. g.][]{Forman1985,O'Sullivan2003,AndersonBregman2011,Dai2012}. With reference to the works by~\citet{MillerBregman2013, MillerBregman2015}, a flattened modification of the $\beta$-model accounting for the disc-like shape of the gas distribution is not considered, since it was conclusively shown to give no sufficient improvement over the spherical profile, and the substructure is not considered in the current analysis. The $\beta$-profile can be written as 
\begin{equation}
    n_e (r) = n_0 (1 + (r/r_c)^2 )^{-3\beta /2},
    \label{eq:beta-profile}
\end{equation}
where $n_e (r)$ is the electron density at the galactocentric radius $r$, $n_0$ is the central density, $r_c$ is the core radius and $\beta$ is the slope parameter. 

The core radius value is considered to be less than $\simeq 5$~kpc, while the data sample does not include any contribution from the region within the radius of $\simeq 6$~kpc. Moreover, this analysis aims to estimate the density at much larger galactocentric distances, $r \gg r_c$, where the halo-associated spherical component is the dominant. Therefore, it is reasonable to rewrite~(\ref{eq:beta-profile}) in a simplified form: 
\begin{equation}
    n_e (r) \simeq (n_0 r_c^{3\beta}) r^{-3\beta} \equiv \alpha r^{-3\beta},
    \label{eq:beta-profile-simplified}
\end{equation}
where $\alpha \equiv n_0 r_c^{3\beta}$ is the normalization. The electron density is thus described by two free parameters, $\alpha$ and $\beta$.

\subsubsection{Metallicity and ion fraction}
Spectroscopic observations provide the information about the total line intensity and number of the absorbing ions along the given line of sight, which is $\propto \int n^2_e Z$ and $\int n_e Z$, respectively. Hence, when estimating the halo density from these observations there is no possibility to directly estimate the shape of the metallicity profile, and this shape must be assumed.

The benefit of the current analysis is that, in contrast to most of previous studies, a common-used constant-metallicity assumption is relaxed. Attempts to parametrize the profile assuming a power law without any physical motivation are also rejected in this work. Instead, a metallicity profile is semi-qualitatively derived from physical considerations. 

Note that the following reasoning is however rather speculative. It provides only qualitative and very simplified physical motivation for the profile. In other words, the latter cannot be claimed neither optimal nor completely inappropriate. None the less, the use of the profile provides a significant improvement over the pre-assumed constant chemical composition consideration, at least, by the very possibility not to assume but to estimate the metallicity from the observational data, and to probe the chemical composition radial dependence. Once this caution has been stated, the profile can be derived as follows. 

Conditionally, the halo gas can be divided into two components, a primordial, almost entirely consisting of hydrogen and helium (i.~e.~$Z\sim 0$), and a late, containing a noticeable amount of metals, i. e. elements that are heavier than helium ($Z\sim Z_{\sun}$). Stars, the producers of the late component, are mainly located in the Galactic disc, near the central region of the halo. Therefore, it is reasonable to suggest the fraction of the late component to decrease with a galactocentric distance. 

For simplicity, it can be assumed that the individual chemical composition of each component is constant. Under this assumption, the metallicity of the gas follows the same radial profile as that of the distribution of the late component fraction. 

It is also assumed that the late gas is mostly held at small galactocentric radii by the gravitational field of the Galaxy, while the primordial gas is rather diffused and volume-filling. 

According to these assumptions, the metallicity profile should be determined by the Galactic gravitational potential as follows: 
\begin{equation}
    Z (r) \propto \exp(-\Phi(r)/\Phi_0),
    \label{eq:metallicity_propto}
\end{equation}
where $\Phi(r)$ is the gravitational potential of the MW and $\Phi_0$ is a scaling constant. The former is predominantly determined by the dark matter density distribution commonly described using the NFW-profile presented in~\citet*{NavarroFrenkWhite1996}:
\begin{equation}
    \rho (r) = \frac{\rho_0}{r/r_0\;(1 + r/r_0)^2},
\end{equation}
where $\rho (r)$ is the dark matter mass density at the given radius, $\rho_0$ is the normalization constant, and $r_0$ is the scale radius. The corresponding potential is
\begin{equation}
    \Phi (r) \propto - \left(r/r_0\right)^{-1} \ln\left(1+r/r_0\right).
    \label{eq:gravpotential_propto}
\end{equation}
Combining~(\ref{eq:gravpotential_propto}) with~(\ref{eq:metallicity_propto}), the metallicity profile can be written as:
\begin{equation}
    Z(r) = a\left(1+r/r_0\right)^{d/r},
    \label{eq:metallicityprofile}
\end{equation}
where $a$ and $d$ are free parameters that are respectively responsible for the metallicity at large galactocentric distances $r \gg d$ and for the slope of the profile. To be consistent with recent estimates \citep[][]{LinLi2019,Ablimit2020,Sofue2020}, the scale radius of the NFW-profile is hereafter fixed at $r_0 = 10$ kpc. 

The \ion{O}{vii} ion fraction is assumed a constant of $f=0.5$. In terms of the fitting procedure, this implies that the metallicity profile is also responsible for the deviations from the true value of this parameter.

\subsubsection{Temperature and optical assumptions}
Following the work by~\citet{MillerBregman2015}, the present model assumes the plasma of the MW halo to be isothermal with the temperature of $\log(T_{\text{halo}}/\text{K}) = 6.3$. The inferred \ion{O}{vii} line emissivity (defined as a total number of radiative transitions per unit volume divided by the square of the electron density) is $\epsilon_{\text{halo}} = 6.05 \times 10^{-15}$ photons cm$^{3}$ s$^{-1}$ \citep[according to][]{Smith2001}. For the further analysis, it is important to emphasize that this emissivity assumes a solar oxygen abundance of $\log(N_{\text{O}}) = 8.93$ from the work by~\citet{AndersGreevesse1989}. 

\citet{MillerBregman2015} found the optical depth corrections to the oxygen emission model to be not very statistically significant, i. e. the results with and without these corrections were consistent within the uncertainty ($\la10$ per cent). These corrections however imply the use of a complicated iterative process. Thus, the current analysis neglects the optical depth effect when modelling the oxygen emission and considers the halo plasma to be optically thin.

Note that the optical corrections are pre-taken into account in the absorption lines data by fitting or assuming Doppler widths. However, these corrections have a noticeably more significant affect on the results than that of the discussed optical depth corrections \citep[][]{MillerBregman2013}, and thus such a consideration is quite reasonable.

The filtering procedure (Subsection~\ref{subsubsec:filtering}) allows to neglect the \ion{H}{i} absorption in the Galactic disc when modelling the oxygen emission. At the considered region, the \ion{H}{i} column density does not exceed~$10^{(18...19)}$~cm$^{-2}$. Thus, the corresponding relative corrections of the line intensities \citep[see][Equation~15]{MillerBregman2015} are $\sim 10^{-(3...4)}$, which is beyond the accuracy of the current analysis. 

\subsubsection{Local Bubble}
When analysing the spectral lines, it is necessary to subtract the contributions that are not due to the MW halo gas. Here, following the work of~\citet{MillerBregman2015} the LB is assumed to be a dominant source of this contamination and all the emission and absorption which is not associated with the CGM is also attributed to the LB. 

It is important to emphasize that this research does not attempt to provide a correct representation of the LB physical properties. The aim is to properly estimate and subtract the contribution from non-signal sources in order to correctly describe the physical properties of the MW halo. For this purpose, a following simple parameterization of the LB is used. 

The LB is assumed a constant-density plasma filling a volume with a radius of $L = 100$ pc \citep[see e. g.][]{Lallement2003,Puspitarinietal2014}. To avoid exceeding the accuracy of the present analysis, the irregular geometry of the LB is neglected, and the LB path length along every sight line is assumed the same. The plasma is considered isothermal with a temperature of $\log(T_{\text{LB}}/\text{K}) = 6.1$ and the inferred \ion{O}{vii} emissivity of $\epsilon_{\text{LB}} = 1.94 \times 10^{-15}$ photons s$^{-1}$ cm$^{3}$ \citep[assuming the solar chemical composition, see][]{Smith2001}.

The LB contribution to the spectroscopic sample is determined by the only free parameter $n_{\text{LB}}$, which characterizes the LB density (in terms of the oxygen spectral lines rather than physical density). 

\subsection{Model fitting}\label{subsec:modelfitting}
\subsubsection{Calculations}
To simplify the fitting procedure, all the errors in the sample are hereafter symmetrized as follows:
\begin{equation}\label{eq:uncertainties}
    \sigma = \left(\frac{\sigma_{+}^{-2} + \sigma_{-}^{-2}}{2}\right)^{-1/2},
\end{equation}
where $\sigma$ is the symmetrized uncertainty, and $\sigma_{\pm}$ are the original uncertainties (with the corresponding sign). This procedure averages the asymmetric errors contribution in terms of chi-squared test.

\paragraph*{Ram-pressure stripping.}
When modelling the ram-pressure stripping data, the observational constraints on the electron density are used as the expected densities at the corresponding galactocentric distances (see Table~\ref{tab:RamPressure}). For the further statistical analysis, the following function is defined:
\begin{equation}
    \chi^2_{\text{dwf}} (\alpha, \beta) = \sum_{i} \left[\frac{(n_e(r^i_{\text{peri}}|\alpha, \beta) - n_e^i)^2}{(\sigma^i_n)^2 + \left(\frac{\upartial n_e (r|\alpha, \beta)}{\upartial r}\bigg|_{r^i_{\text{peri}}}\cdot\sigma^i_r\right)^2}\right],
    \label{eq:chi2dwf}
\end{equation}
where $i$ is the integer index numbering the observations in Table~\ref{tab:RamPressure}, $n_e^i$~and~$r_{\text{peri}}^i$~are the density and radius estimates, respectively, $\sigma^i_{n,r}$~are~the corresponding uncertainties (symmetrized according to Equation~\ref{eq:uncertainties}), and $n_e(r|\alpha,\beta)$ is the parametric electron density profile defined by Equation~(\ref{eq:beta-profile-simplified}). The latter notation is hereafter used to highlight, where needed, the parametric representation of a function.

\paragraph*{Oxygen spectra.}
The spectroscopic observations probe the gas properties along the individual lines of sight, while the parametric profiles describing the halo are radial. To take into account the Sun displacement from the Galactic centre, the following coordinate transformation is used:
\begin{equation}
    r^2 = r^2_{\sun} + s^2 - 2\,s\,r_{\sun}\cos(b)\cos(l), 
    \label{eq:r-s}
\end{equation}
where $s$ is the sight line coordinate, $l$ and $b$ are the galactic longitude and latitude, respectively, and $r_{\sun}$ is the distance between the Sun and the MW centre (hereafter, $r_{\sun} = 8.5$ kpc is assumed). 

The cutoff coordinate ($r(s_{\text{cut}}) = R_{\text{vir}}$) is determined as:
\begin{equation}
    s_{\text{cut}} = r_{\sun}\left(\cos(b)\cos(l)+\sqrt{\cos(b)^2\cos(l)^2+R_{\text{vir}}^2/r_{\sun}^2-1}\right)
\end{equation}

\subparagraph*{\ion{O}{vii} absorption.}
To model the \ion{O}{vii} halo column density, the ion density should be integrated along the line of sight as follows:
\begin{equation}
    N_{\text{halo}} = 7.0 \times 10^{13} \left[\int\limits_0^{s_{\text{cut}}} \frac{n_e (r(s))}{10^{-4} \text{ cm}^{-3}}\frac{Z(r(s))}{Z_{\sun}}\,\frac{ds}{1\text{ kpc}}\right]\;\text{cm}^{-2},
    \label{eq:columndensityhalo}
\end{equation}
where the electron density $n_e (r)$, the metallicity $Z(r)$ and the radius $r(s)$ are respectively determined by Equations~(\ref{eq:beta-profile-simplified}). (\ref{eq:metallicityprofile}) and (\ref{eq:r-s}). 

It should be however kept in mind that the ion fraction of $f=0.5$ is assumed. Thus, $Z(r)$ in fact represents both the true metallicity and ion fraction as $Z_{\text{true}}(r)f_{\text{true}}(r)/0.5$. The same remark applies to the oxygen emission lines.

The LB also has a contribution to the column density:
\begin{equation}
    \begin{gathered}
    N_{\text{LB}} = 7.0 \times 10^{13} \left(\frac{n_{\text{LB}}}{10^{-3}\;\text{cm}^{-3}}\right)\left(\frac{L}{100\;\text{pc}}\right)\;\text{cm}^{-2},
    \end{gathered}
    \label{eq:columndensityLB}
\end{equation}
where $n_{\text{LB}}$ and $L = 100$ pc are respectively the density parameter and the path length of the LB discussed in Subsection~\ref{subsec:modeldescription}. 

The total column density $N_{\text{\ion{O}{vii}}}$ is the sum of the halo and the LB contributions (see Equations~\ref{eq:columndensityhalo} and \ref{eq:columndensityLB}, respectively):
\begin{equation}
    N_{\text{\ion{O}{vii}}} = N_{\text{halo}} + N_{\text{LB}}
    \label{eq:columndensity}
\end{equation}

For the statistical analysis, the following function is then defined: 
\begin{equation}\label{eq:chi2abs}
    \chi^2_{\text{abs}} (\alpha, \beta, a, d, n_{\text{LB}})
    = \sum_{i} \left[\frac{\left(N_{\text{\ion{O}{vii}}} (l^i, b^i|\alpha, ...) -  N^i_{\text{\ion{O}{vii}}}\right)^2}{(\sigma^i_N)^2}\right],
\end{equation}
where $i$ is the integer index numbering the observations (and the corresponding sight lines), $N^i_{\text{\ion{O}{vii}}}$ is the column density in the direction with the galactic longitude and latitude of~$l^i$~and~$b^i$, respectively, inferred from the observations, $\sigma_N^i$ is the corresponding symmetrized uncertainty, and $N_{\text{\ion{O}{vii}}} (l^i, b^i|\alpha,...) \equiv N_{\text{\ion{O}{vii}}} (l^i, b^i|\alpha, \beta, a, d, n_{\text{LB}})$ is the modelled oxygen column density determined by Equation~(\ref{eq:columndensity}).

\subparagraph*{\ion{O}{vii} emission.}
To model the \ion{O}{vii} line strengths assuming an optically thin plasma, the halo intensity is calculated as follows: 
\begin{equation}
    \frac{dI_{\text{halo}}}{ds} \equiv j(s) = \frac{\epsilon_{\text{halo}}}{4\pi}\left(n_e^2 (r) \frac{Z(r)}{Z_{\sun}}\right)\bigg|_{r=r(s)}
    \label{eq:dI/ds}
\end{equation}
In this equation, the correction factor of $Z(r)/Z_{\sun}$ is reasoned by the fact that $j(s) \propto n_{\text{ion}} n_e \propto n_e^2 Z$. The halo emissivity $\epsilon_{\text{halo}}$ assumes the solar oxygen abundance (Subsection \ref{subsec:modeldescription}), while in the present model the true amount of ions decreases with radius. The latter is thus taken into account.

Integrating Equation~(\ref{eq:dI/ds}) along the line of sight, the following expression for the halo contribution to the line strength is obtained:
\begin{equation}\label{eq:Ihalo}
    I_{\text{halo}} = 1.5\times10^{-2}\left[\int\limits_0^{s_{\text{cut}}} \left(\frac{n_e (r(s))}{10^{-4}\text{ cm}^{-3}}\right)^2 \frac{Z(r(s))}{Z_{\sun}}\,\frac{ds}{1\text{ kpc}}\right]\; \text{L. U.} 
\end{equation}

The LB contribution is calculated as follows: 
\begin{equation}\label{eq:ILB}
    I_{\text{LB}} = \frac{\epsilon_{\text{LB}}}{4\pi}\left(n^2_{\text{LB}}L\right) = 4.75 \times 10^{-2} \left(\frac{n_{\text{LB}}}{10^{-3}\;\text{cm}^{-3}}\right)^2\;\text{L. U.},
\end{equation}
where $\epsilon_{\text{LB}}$ is the emissivity of the LB discussed in Subsection~\ref{subsec:modeldescription}.

The total \ion{O}{vii} emission line intensity is the sum of the halo and the LB contributions (Equations~\ref{eq:Ihalo} and \ref{eq:ILB}, respectively):
\begin{equation}\label{eq:intensity}
    I_{\text{\ion{O}{vii}}} = I_{\text{halo}} + I_{\text{LB}}
\end{equation}

Once the uncertainties are symmetrized, a systematic uncertainty (Subsection~\ref{subsec:Oviispectra}) is added in quadrature: 
\begin{equation}\label{eq:addeduncertainty}
    \sigma_I = \left(\left(\sigma^{\text{sym}}_I\right)^2 +    \left(\sigma^{\text{add}}_I\right)^2\right)^{1/2},
\end{equation}
where $\sigma^{\text{add}}_I$ is the additional uncertainty, $\sigma^{\text{sym}}_I$ is the symmetrized uncertainty, and $\sigma_I$ is the resulting uncertainty.

The function for the further statistical analysis is defined as: 
\begin{equation}\label{eq:chi2emis}
    \chi^2_{\text{emis}} (\alpha, \beta, a, d, n_{\text{LB}}) = \sum_{i} \left[\frac{\left(I_{\text{\ion{O}{vii}}} (l^i, b^i|\alpha,...) -  I^i_{\text{\ion{O}{vii}}}\right)^2}{(\sigma^i_I)^2}\right],
\end{equation}
where $i$ is the integer index numbering the observations (and corresponding sight lines), $I^i_{\text{\ion{O}{vii}}}$ is the measured intensity in the direction with the galactic longitude and latitude of~$l^i$~and~$b^i$, respectively, $\sigma_I^i$ is the corresponding uncertainty (see Equation~\ref{eq:addeduncertainty}), and $I_{\text{\ion{O}{vii}}} (l^i, b^i|\alpha,...) \equiv I_{\text{\ion{O}{vii}}} (l^i, b^i|\alpha, \beta, a, d, n_{\text{LB}})$ is the modelled intensity in the direction of $(l^i, b^i)$ determined by Equation~(\ref{eq:intensity}).

\subparagraph*{}
The statistical function accounting for the entire oxygen spectra data is defined as:
\begin{equation}\label{eq:chi2oxy}
    \chi^2_{\text{oxg}} (\alpha, \beta, a, d, n_{\text{LB}}) =  \chi^2_{\text{abs}} (\alpha,...) +  \chi^2_{\text{emis}} (\alpha,...),
\end{equation}
where $(\alpha, ...)\equiv(\alpha, \beta, a, d, n_{\text{LB}})$, and $\chi^2_{\text{abs}}$, $\chi^2_{\text{emis}}$ are defined by Equations~(\ref{eq:chi2abs}) and~(\ref{eq:chi2emis}), respectively. The additional uncertainty (Equation~\ref{eq:addeduncertainty}) is increased until the acceptable~$\chi^2_{\text{oxg}}$ is found (this will be discussed below). In this work, the required additional uncertainty of $\sigma^{\text{add}}_I = 2.37$ L. U. is concluded.

\paragraph*{Likelihood definition.}
This study aims to consistently describe the observations of ram-pressure stripping and oxygen spectra. In terms of the fitting procedure this means to find the optimal, physically meaningful set of parameters $(\alpha^*, \beta^*, a^*, d^*, n^*_{\text{LB}})$ that effectively minimizes $\chi^2_{\text{dwf}}$ (Equation~\ref{eq:chi2dwf}) as well as  $\chi^2_{\text{oxg}}$ (Equation~\ref{eq:chi2oxy}). Thus, the goal is to explore the 5-dimensional parameter space in order to minimize the following function:
\begin{equation}\label{eq:chi2total}
    \chi^2 (\alpha, \beta, a, d, n_{\text{LB}}) = \frac{w_{\text{dwf}}}{2}\cdot\chi^2_{\text{dwf}} (\alpha, \beta) +  \frac{w_{\text{oxg}}}{2}\cdot\chi^2_{\text{oxg}} (\alpha, \beta, a, d, n_{\text{LB}})
\end{equation}
In this equation, $w_{\text{dwf}}$ and $w_{\text{oxg}}$ are the positive constants representing statistical weight of the corresponding data sub-sample. 

Although the ram-pressure stripping sample consists of only 7 observations (while the spectral one includes 449), it has an essential advantage, since it constrains the density profile independently of the metallicity profile and the LB density. The use of the unweighted sum of the two functions, $\chi^2_{\text{dwf}}+\chi^2_{\text{oxgn}}$, would suppress this substantial contribution, and its affect on the result would become negligible. This, in turn, would reduce the quality of the further analysis, and thus the weighted sum is used in order to balance the two contributions. 

The weights in Equation~(\ref{eq:chi2total}) are defined as follows:
\begin{equation}
    w = \left(\text{ICDF} (p\text{-value}=0.5, \text{dof} = \text{number of observations})\right)^{-1},
\end{equation}
where ICDF ($p$-value, dof) is the inverse cumulative distribution function of the $\chi^2_{\text{dof}}$ distribution. This definition implies the desirable $\chi^2 (\alpha^*, \beta^*, a^*, d^*, n^*_{\text{LB}}) \simeq 1/2 + 1/2 = 1$ (note that $\chi^2$ actually represents $\chi^2/\text{dof}$, but "$/\text{dof}$" is hereafter generally omitted for brevity). The likelihood function is defined as $\mathcal{L}=\exp(-\chi^2/2)$.

\subsubsection{Fitting procedure}\label{subsubsec:fitting}
Fitting procedure aims not just to find a maximum of the likelihood, but to maximize the latter by finding the physically meaningful model parameter set optimally describing the properties of the halo. 

Herewith, the 5-dimensional parameter space is associated with three parameters ($a$, $d$ and $n_{\text{LB}}$) that are poorly constrained by observations. Hence, it was unreasonable to impose strict 5-dimensional constraints. Instead, a step-by-step approach was chosen.

\paragraph*{Step 1.}
The first step in the fitting process was to estimate the metallicity profile parameters that could describe the data under consideration consistently the supplementary electron density observational constraints (Subsection~\ref{subsec:supplementarydata}).

For this purpose, the LB density parameter of $\bar n_{\text{LB}} = 4.0\times10^{-3}$~cm$^{-3}$ was fixed \citep[the best-fitting parameter from the work of][]{MillerBregman2015}, and $\chi^2$ was then averaged over the $(\alpha, \beta)$-area defined by the electron density constraints discussed in Subsection~\ref{subsec:supplementarydata}:
\begin{equation}
    \Omega = (\alpha, \beta):~\begin{cases}
    \int\limits_{0}^{223\text{ kpc}} \frac{dr 3 r^2}{(223\text{ kpc})^{3}} \frac{n_e(r|\alpha, \beta)}{10^{-4}\text{ cm}^{-3}} > 0.13 \\
    \int\limits_{0}^{50\text{ kpc}} \frac{ds}{50\text{ kpc}}\frac{n_e(r|\alpha, \beta)}{10^{-4}\text{ cm}^{-3}}\bigg|_{r=r(s)}< 5
    \end{cases},
    \label{eq:Omega}
\end{equation}
where $n_e(r|\alpha, \beta)$ is from Equation~(\ref{eq:beta-profile-simplified}), and $r(s)$ is defined by Equation~(\ref{eq:r-s}) for the LMC coordinates, $(l, b)=(280\degr,-35\degr$). Then:
\begin{equation}
    \begin{gathered}
    \chi^2_{\text{St1}} (a, d) = w_{\text{oxg}} \cdot \langle\chi^2_{\text{oxg}} (\alpha, \beta, a, d, n_{\text{LB}} = \bar n_{\text{LB}})\rangle_{(\alpha, \beta)} \equiv \\
    \equiv w_{\text{oxg}} \cdot \left(\iint\limits_{\Omega} d\alpha d\beta\,\chi^2_{\text{oxg}} (\alpha, \beta, a, d, n_{\text{LB}} = \bar n_{\text{LB}})\right)\cdot\left(\iint\limits_{\Omega} d\alpha d\beta \right)^{-1}
    \end{gathered}
    \label{eq:chi2step1}
\end{equation}
Once function~(\ref{eq:chi2step1}) is defined, it should be minimized in order to obtain the metallicity profile parameters. Herewith, physically reasonable constraints of $a < Z_{\sun}$ and $ a, d > 0$ are imposed, and an initial guess of $a = 0.3~Z_{\sun}$, $d = 0$ (the constant-metallicity profile $Z=0.3~Z_{\sun}$ assumed in previous studies) is used. The parameters minimizing $\chi^2_{\text{St1}}$ are hereafter referred to as $\bar a, \bar d$.

\paragraph*{Step 2.} 
After finding $\bar a$ and $\bar d$, the respective parameters in $\chi^2 (\alpha, ...)$ were fixed (Equation~\ref{eq:chi2total}):
\begin{equation}
    \chi^2_{\text{St2}} (\alpha, \beta) = \chi^2 (\alpha, \beta, a = \bar a, d = \bar d, n_{\text{LB}} = \bar n_{\text{LB}})
    \label{eq:chi2step2}
\end{equation}
The second step aimed to find the $\beta$-profile parameters that are suitable for the temporarily fixed metallicity profile and do not substantially disagree with the constraints discussed at the previous step. For this purpose, constraints of $\log\left(\alpha/10^{-4}\text{ cm}^3\text{kpc}^{3\beta}\right) < 1.5$, $\beta < 1$ and $\alpha, \beta > 0$ were imposed, and then 2-dimensional parameter space was explored to find a minimum of function~(\ref{eq:chi2step2}). The parameters minimizing $\chi^2_{\text{St2}}$ are hereafter referred to as $\bar \alpha, \bar \beta$.

\paragraph*{Step 3.} 
Finally, $(\bar \alpha, \bar \beta, \bar a, \bar d, \bar n_{\text{LB}})$ was interpreted as an acceptable parameter set approximating the optimal one and used as an initial guess when exploring the 5-dimensional parameter space and finding a minimum of the total $\chi^2 (\alpha, ...)$:
\begin{equation}
    \begin{gathered}
    \chi^2_{\text{St3}} (\alpha, \beta, a, d, n_{\text{LB}}) = \chi^2 (\alpha, \beta, a, d, n_{\text{LB}}), \\
    \text{initial guess} =  (\bar \alpha, \bar \beta, \bar a, \bar d, \bar n_{\text{LB}})
    \end{gathered}
\end{equation}
\paragraph*{Step 4.} 
Once the best-fitting parameters were found, it was necessary to estimate the corresponding uncertainties. For this purpose, discrete values of $\mathcal{L}$ were calculated in a best-fitting parameter $x$ neighbourhood with the other parameters fixed, and $\sigma^{\pm}_{x}$ corresponding to 68 per cent CL of $x$ were determined by $\log\left(\mathcal{L}(x)/\mathcal{L}(x\pm\sigma^{\pm}_{x})\right) = 0.5 \Rightarrow \Delta \chi^2 = 1$. By the similar approach, a joint posterior probability distribution of $\beta$-profile parameters were obtained. 

\begin{table}
	\centering
	\caption{Step-by-step fitting results. The best-fitting parameters correspond to $\chi^2\text{ (dof)}\simeq1.43\,(458)$, but then $\chi^2_{\text{true}} \simeq 1.02$ is argued for (see Subsection~\ref{subsec:selfconsistency}). In Sub-subsection~\ref{subsubsec:fitting}, a detailed discussion of the procedure is presented.}
	\label{tab:Results}
    \begin{tabular}{l c c c c c}
        \hline 
        \hline
         $\alpha$, cm$^{-3}$ kpc$^{3\beta}$ & $\beta$ & $a$, $Z_{\sun}$ & $d$, kpc & $n_{\text{LB}}$, cm$^{-3}$\\
        \hline
         6.58$^{+1.7}_{-2.1}\times10^{-3}$ & 0.337$^{+0.043}_{-0.028}$ & 0.29$^{+0.18}_{-0.18}$ & 8.5$^{+7.6}_{-15.9}$ & 4$^{+6}_{-13}\times10^{-3}$ \\
        \hline
        \end{tabular}
\end{table}

\begin{figure}
	\includegraphics[width=\columnwidth]{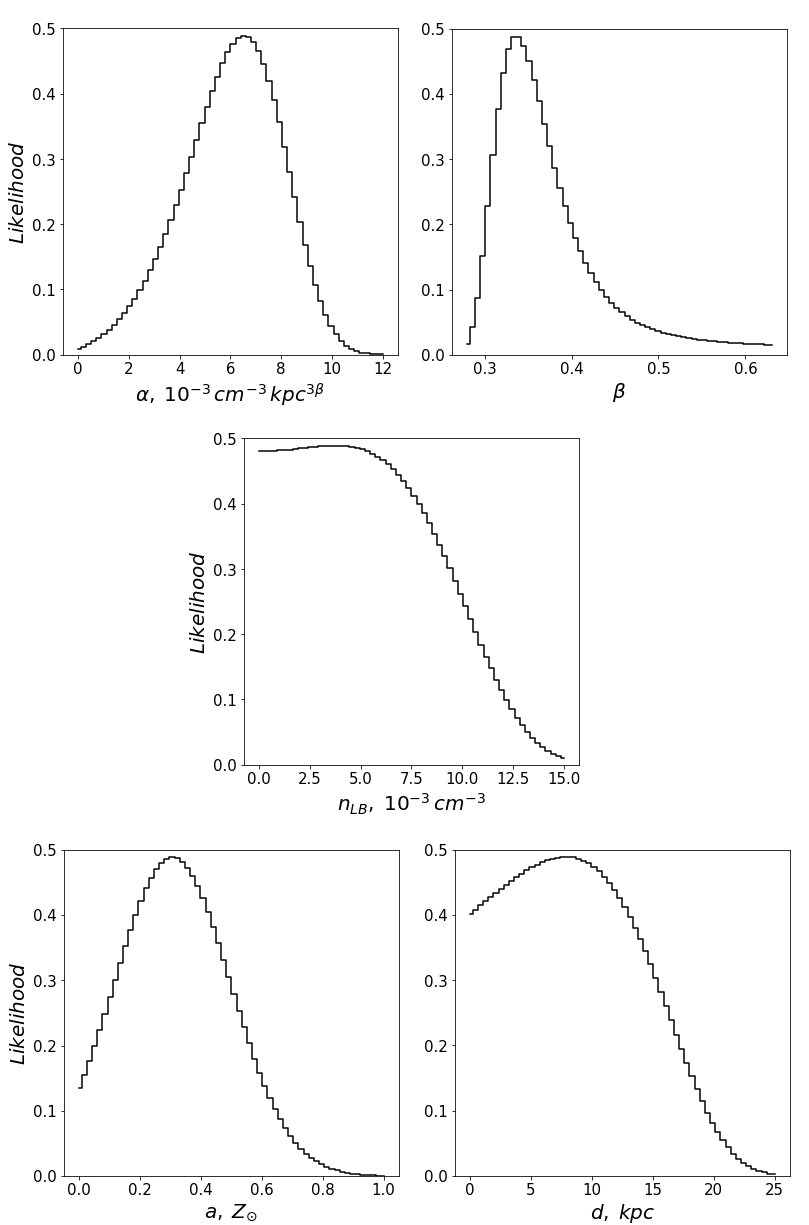}
    \caption{Likelihood function as a function of individual parameters with the other parameters fixed at their best-fitting values. See also  Table~\ref{tab:Results}.}
    \label{fig:likelihood}
\end{figure}

\begin{figure}
	\includegraphics[width=\columnwidth]{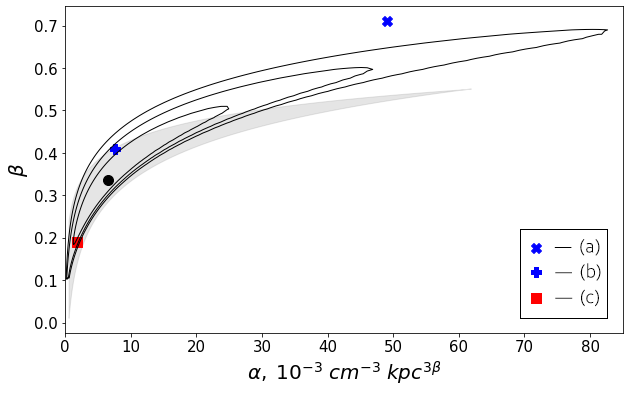}
    \caption{Joint posterior probability distribution for $\beta$-profile parameters. The contours represent $1\sigma$, $2\sigma$ and $3\sigma$ confidence regions. The points represent the best-fitting parameters obtained by this work (black circle), (a) \citet{MillerBregman2013}, (b) \citet{MillerBregman2015}, and (c) \citet{Troitsky2017}. The shaded area corresponds to the observational constraints \citep[][]{BlitzRobishaw2000,AndersonBregman2010}.}
    \label{fig:betaparameters}
\end{figure}

\subsection{Results}\label{subsec:results}
The results of the step-by-step fitting procedure are briefly presented in Table~\ref{tab:Results} and Figures~\ref{fig:likelihood}, \ref{fig:betaparameters}. Section~\ref{sec:discussion} presents a detailed discussion.

\section{Discussion}\label{sec:discussion}
This section discusses self-consistency, limitations and stability of the model, provides physical implications for the MW and compares the results with previous studies.

Since the physical quantities to be calculated may depend on more than one of the model parameters, their relative uncertainties are hereafter estimated as follows:
\begin{equation}
    \sigma_{Y}/Y = \sqrt{\sum^{N}_{i=1}\left(\sigma^{i}_{Y}/Y\right)^2},
\end{equation}
where $Y$ is the physical quantity depending on $N$ model parameters~$x_{i}$, $i$ is the integer index numbering the parameters, $\sigma^{i}_{Y}/Y$ is the relative uncertainty corresponding to the uncertainty of $x_{i}$.

Whenever the calculated uncertainty implies a negative lower bound for a positive physical parameter, this  is remained unchanged in order to emphasize that the current analysis is unable to statistically rule out physically meaningless values of this parameter (see e.~g.~Table~\ref{tab:Results}, where the metallicity profile slope parameter $d>0$ is estimated as $d=8.5^{+7.6}_{-15.9}\text{ kpc}$). This allows to quantitatively demonstrate the actual quality of the analysis.

\subsection{Electron density \& metallicity}\label{subsec:electrondensity_metallicity}
\subsubsection{Electron density}
The halo-associated electron density profile behaviour can be approximately described as $\propto r^{-1}$, which is considerably flatter than that presented in \citet{MillerBregman2013,MillerBregman2015}, where $r^{-(1.5...2)}$ was obtained, but more sharp than $r^{-(0.45...0.75)}$, which was reported by \citet{Troitsky2017}, and is relatively close to that of~\citet{Kaaretetal2020}, where $r^{-(1.11...1.23)}$ is obtained. The profile obtained here is also slightly flatter (but consistent within $2\sigma$) than that reported in~\citet{LiBregman2018}. The authors obtained the best-fitting slope of $r^{-(1.16...1.22)}$ by modelling the X-ray profiles of six isolated nearby ($z \la 0.02$) massive (with the stellar mass of $\ga 1.5\times 10^{11}~\text{M}_{\sun}$) spiral galaxies. The slope obtained here is also consistent (within $(1...2)\sigma$) with that obtained for other massive spiral galaxies, see~e.~g.~\citet{Bogdan2013}, where best-fitting profile behaviour of $\propto r^{-0.87}$ and $r^{-1.11}$ were reported for NGC~1961 and NGC~6753, respectively. The profile obtained here is however flatter than typical density distribution around an external galaxy for which metallicity can be measured.

Not surprisingly, the profile has a flatter slope than that obtained in the most of earlier studies of the MW based on spectroscopic data. Previously, the gas metallicity was commonly assumed a constant. Hence, in terms of the $\beta$-profile slope parameter, $\beta=\beta_{\text{true}}+\beta_{\text{add}}$ was constrained in fact, where $\beta_{\text{true}}$ is the actual electron density profile slope parameter, and $\beta_{\text{add}}$ is the contribution from the gas metallicity and ion fraction gradient which depends on the type of spectroscopic observations. In the current analysis, the model accounts for the negative metallicity gradient and thus gives a more appropriate constraint for $\beta$ without the contribution of $\beta_{\text{add}}$, which obviously corresponds to a flatter electron density profile \citep[see][]{Troitsky2017,BregmanAnderson2018}.

It is notable that the resulting electron density profile is flat not due to the ram-pressure stripping constraints, since for the spectroscopic data separately, the best-fitting slope is $r^{-0.92}$. When fixing metallicity at $Z=0.3~Z_{\sun}$, the slope of $r^{-1.1}$ is obtained, reproducing the result of $r^{-(1.1...1.3)}$ reported by~\citet{MillerBregman2015} based on the same \ion{O}{vii} emission sample (the minor flattening is likely to be due to the excluding the region of $|b|<30\degr$ in the present study, while in the previous work the authors excluded only the region of $|b|<10\degr$). 

The best-fitting profile, however, is consistent within $1\sigma$ with that obtained in the work by~\citet{MillerBregman2015}.

\begin{figure}
	\includegraphics[width=\columnwidth]{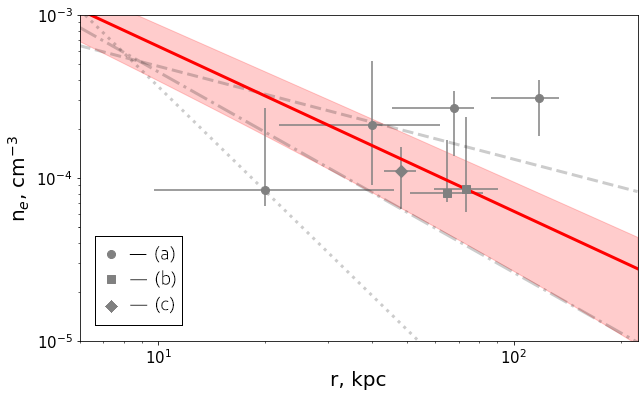}
    \caption{Electron density profiles of the circumgalactic gas. The red line and the red-shaded area represent the best-fitting profile and $1\sigma$ allowed range obtained in this work, respectively. The gray lines represent the best-fitting profiles obtained in the works by~\citet{MillerBregman2013} (dotted), \citet{MillerBregman2015} (dash-dotted), and \citet{Troitsky2017} (dashed). The points with error bars represent the observational constraints obtained by (a) \citet{GrcevichPutman2009}, (b) \citet{Gattoetal2013}, and (c) \citet{Salemetal2015}.}
    \label{fig:betaprofile}
\end{figure}

Figure~\ref{fig:betaprofile} compares the best-fitting profile with those obtained in previous studies and with the ram-pressure stripping constraints that were considered in this work.

It can be concluded that in the inner part of the halo ($r\la60$ kpc), $n_e \sim 10^{-4}$ cm$^{-3}$, while in the outer part ($r\ga60$ kpc) $n_e \sim 10^{-5}$ cm$^{-3}$, which generally agrees with the ram-pressure stripping data. A more detailed self-consistency examination is provided in Subsection~\ref{subsec:selfconsistency}.

\subsubsection{Metallicity}
The best-fitting metallicity profile is rather flat, decreasing from $\simeq 0.5~Z_{\sun}$ at near-solar galactocentric radii to $\simeq 0.3~Z_{\sun}$ in the outer part of the MW halo, which is well consistent with the constraints presented by \citet{MillerBregman2013, MillerBregman2015}, but higher than the result reported by \citet{Troitsky2017} by the factor of $\simeq 2$.

This study also reflects the results of~\citet{Voit2019}, where the metallicity is reported to decrease from $\simeq Z_{\sun}$ at $r\sim 10$ kpc to $\simeq 0.3~Z_{\sun}$ at $r\sim 200$ kpc. All of these metallicity models are consistent with that obtained here within $\simeq1\sigma$. Also, the slope of the profile obtained here accords well with that estimated by~\citet{MillerBregman2015}, where $Z\propto r^{-0.2}$ is derived by dividing the electron density profiles obtained separately for absorption and emission spectroscopic data samples under the constant-metallicity assumption.

However, due to the large uncertainty of the parameters, this work cannot definitively conclude whether or not the metallicity gradient is sharp or flat. The gas metallicity is only constrained as $\simeq(0.1...0.7)~Z_{\sun}$ at $r\ga50$ kpc (assuming this particular shape of the metallicity profile, see Subsection~\ref{subsec:finalremarks} for a discussion).

\begin{figure}
	\includegraphics[width=\columnwidth]{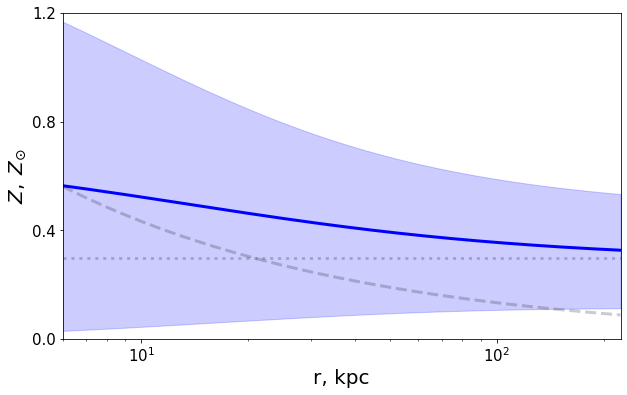}
    \caption{Metallicity profiles of the circumgalactic gas. The blue line and the blue-shaded region represent the best-fitting profile and the $1\sigma$ allowed region obtained in this work, respectively. The gray lines represent the constant-metallicity profile assumed in previous studies (dotted) and the best-fitting profile obtained in the work by~\citet{Troitsky2017} (dashed).}
    \label{fig:metaprofile}
\end{figure}

None the less, such an approach allows to constrain the halo gas metallicity at different galactocentric radii from the spectroscopic observations \citep[note that in the works by][only a lower limit for the metallicity was derived, while the current analysis also provides an upper limit]{MillerBregman2013,MillerBregman2015}. Moreover, by relaxing the constant-metallicity assumption, the discrepancies in modelling of these and other observations can be resolved, which might be quite useful in further studies.

\subsection{Limitations \& stability}\label{subsec:limitations}
Here, the reliability of the results and the main limitations on the model applicability are discussed. 

\subsubsection{Limitations}\label{subsubsec:limitations}
It is important to emphasize that the profiles obtained here cannot be validly extrapolated to all the regions of the halo. They are applicable only at the distances of $6$~kpc~$=R_{\text{in}}<r<R_{\text{vir}}=223$~kpc, since this work did not consider any observational data within the region of $r<6$~kpc, and the value of $223$ kpc was used as a halo cutoff radius. The inner truncation validates the simplified form of the $\beta$-profile~(\ref{eq:beta-profile-simplified}), since $(r/r_c)^2 \gg 1$ in the region of applicability. 

Similarly, observational data from the regions near the Galactic plane and bulge were not considered. Therefore, the profiles are inapplicable beyond the galactic latitudes of $|b| > 30\degr$ and longitudes of $10\degr < l < 350\degr$. Specifically, the true density profile must differ significantly from that constrained by this work in the region near the MW disc $(r\la 20 \text{ kpc})$, where the axial symmetric distribution of the disc-associated gas dominates over the spherical halo-associated profile. 

To be more precise, the profiles applicability beyond the discussed regions is very limited and, if needed, this should be done with a necessary caution, because the quantitative results are highly likely to be distorted.

Besides that, it should be kept in mind that the sample was reduced by excluding the data corresponding to the high-density regions, and thus the profiles describe only the background density and chemical composition of the MW halo completely ignoring its substructure. 

It is also important that the metallicity profile assumed the ionization fraction of $f=0.5$ and, in fact, constraints not the metallicity but $Z_{\text{true}}(r)f_{\text{true}}(r)/0.5$.

\subsubsection{Stability of the fitting procedure}\label{subsubsec:stability}
\paragraph*{LB density.} 
The most poorly-constrained (by both observations and this work) parameter in the considered model is the LB density. Design of the fitting procedure could have generated a somewhat bias, since the initially assumed LB density was fixed when approximating the profiles and considered as a free parameter only at the final step.

This requires to test whether or not the fitting procedure is stable in terms of minor variations of the LB density initial guess. In order to explore this, additional fitting procedures are conducted with the LB contribution neglected ($n_{\text{LB}}=0$), and with the initial guess of $n_{\text{LB}} = \bar n_{\text{LB}} \pm 1\times10^{-3}\text{ cm}^{-3}$, using a step-by-step algorithm similar to the discussed in Sub-subsection~\ref{subsubsec:fitting}.

The results are presented in Table~\ref{tab:lb_stability}. As can be seen, minor variations of the initial LB density parameter ($\simeq 0.25\bar n_{\text{LB}}$) almost do not affect the results, which indicates the fitting procedure to be stable. 

If the LB contribution is neglected, minor changes (within the uncertainty) in both electron density and metallicity profile parameters arise.

None the less, the LB density parameter under discussion is still very poorly constrained, and it is not clear whether this contribution to the spectroscopic observations is truly negligible or this finding is rather due to the large systematic uncertainties of the sample.
\begin{table}
	\centering
	\caption{Results of the model stability test in terms of variations of the initial LB density (see Sub-subsection~\ref{subsubsec:stability}). The top and the bottom lines of the table correspond, respectively, to the fixed value of $n_{\text{LB}}=0$ and to the initial guess of $n_{\text{LB}} = \bar n_{\text{LB}} \pm 1\times10^{-3}\text{ cm}^{-3}$ (since there is no statistical difference between the latter two, the results are shown together).}
	\label{tab:lb_stability}
    \begin{tabular}{l c c c c c}
        \hline 
        \hline
        $\alpha$, cm$^{-3}$ kpc$^{3\beta}$ & $\beta$ & $a$, $Z_{\sun}$ & $d$, kpc & $n_{\text{LB}}$, cm$^{-3}$\\
        \hline
        6.0$^{+1.5}_{-1.8}\times10^{-3}$ & 0.33$^{+0.04}_{-0.03}$ & 0.47$^{+0.26}_{-0.26}$ & 3.7$^{+7.1}_{-13.5}$ & 0\\
        \hline
        6.6$^{+1.7}_{-2.1}\times10^{-3}$ & 0.34$^{+0.04}_{-0.03}$ & 0.29$^{+0.19}_{-0.19}$ & 8.5$^{+7.6}_{-15.9}$ & 4$^{+6}_{-13}\times10^{-3}$ \\
        \hline
        \end{tabular}
\end{table}
\paragraph*{Other parameters.} 
Speculatively, the model is expected to be stable to the variations of (a) $R_{\text{vir}}=223$ kpc \citep[the median Galactic virial radius inferred from \textit{Gaia} DR2 in the work by][]{Wang2020} and (b) $r_0=10$ kpc \citep[the NFW profile scale radius consistent with the recent estimates of][]{LinLi2019,Ablimit2020,Sofue2020} in terms of the resulting density and metallicity at given galactocentric radius. This is argued for by the following: (a) the gas density in the outer region of the MW halo ($r\ga60$ kpc) is by (1...2) orders of magnitude lower than that in the inner region, and therefore minor variations of the halo cutoff radius lead to minor variations of the modelled column densities ($\sim 10^{14}$ cm$^{-2} \times \Delta R/10\text{ kpc} \sim 10^{-3} \times \Delta R/10\text{ kpc}$ of the median observed column density) and emission line intensities ($\sim 10^{-3}$ L. U. $\times \Delta R/10\text{ kpc} \sim 10^{-4} \times \Delta R/10\text{ kpc}$ of the median observed intensity), and, which is notable, do not lead to any variations of the modelled densities at given radius when modelling ram-pressure stripping observations; (b) the variations of $r_0$ in the considered model can easily be compensated by the corresponding variation of $d$ without any noticeable difference in the metallicity at any given radius within the region of applicability. 

The variations of $R_{\text{vir}}$, however, might affect the implied integral physical properties such as the total mass of the gas more significantly (the relative variations can be estimated as $\Delta M/M \sim 3\Delta R/R$). None the less, a detailed quantitative analysis of these effects is beyond the scope of this work and it will not be discussed further. 

\subsection{Self-consistency}\label{subsec:selfconsistency}
To argue for the model self-consistency, this subsection explores the results in terms of the observational data and initial assumptions.

\subsubsection{Observational constraints}\label{subsubsec:rampressure_oxygen}
\paragraph*{Ram-pressure stripping.} 
In Figure~\ref{fig:betaprofile}, ram-pressure stripping electron density constraints compared with the best-fitting density profile are shown. Within the uncertainty, the profile is consistent with 5 out of 7 considered observations, including the most accurate constraints \citep[][]{Gattoetal2013,Salemetal2015}.

\begin{figure}
	\includegraphics[width=\columnwidth]{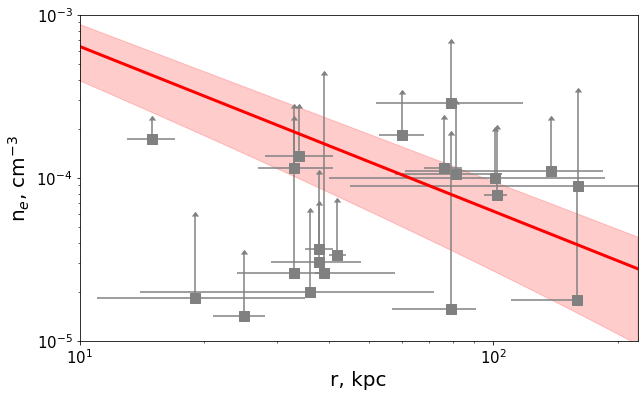}
    \caption{Electron density of the circumgalactic gas obtained in this work and its 68 per cent CL region (red solid line and red-shaded area, respectively) compared with the ram-pressure stripping constraints presented in the work by~\citet{Putmanetal2021} (gray points with error bars), see Sub-subsection~\ref{subsubsec:rampressure_oxygen}.}
    \label{fig:rampressure}
\end{figure}

It seems that the best-fitting $\chi^2\simeq(1.80+1.06)/2=1.43$ (where 1.80 is the ram-pressure stripping contribution) indicates a somewhat under-fit of ram-pressure stripping data. It is however concluded to have been caused by the uncertainties symmetrization which affect these data contribution significantly due to considerably asymmetric uncertainties. To argue for that, $\chi^2$ is recalculated with respect to this asymmetry and $\chi_{\text{true}}^2\simeq (0.98 + 1.06)/2 = 1.02$ is found. 

It is also notable that, with respect to asymmetric uncertainties and excluding the data associated with the Carina dwarf (since the pericentre estimates for this satellite assumed in the simulations are not consistent with the recent observational constraints with the discrepancy of $\simeq(3...4)\sigma$, while the other assumed satellites' pericentres are consistent with their improved values within $1\sigma$), $\chi^2$ value is $\simeq (1.37 + 1.06)/2 = 1.22$. Thus, it can be speculatively stated that the model parameters obtained here would probably describe recent ram-pressure stripping data quite well (once re-simulations would have been provided). 

According to the work of \citet{BlitzRobishaw2000}, the halo electron density averaged over a volume of 250 kpc radius should be $\langle n_e\rangle \ga 1.3\times10^{-5}$ cm$^{-3}$. The volume-averaged (over the region of the results applicability, i. e. from $R_{\text{in}}=6$ kpc up to $R_{\text{vir}}=223$ kpc) electron density inferred by this work is $\langle n_e \rangle \simeq 4.2\,(1.8...6.6)\times10^{-5}$~cm$^{-3}$, which corroborates the result of the previous study. This can be also considered as an argument for the absence of internal contradictions in the current analysis, since the constraint under discussion was used in the fitting procedure.

Also, the resulting electron density profile is in agreement with the constraints presented by~\citet{Putmanetal2021}. The authors estimated the halo density at satellite's pericentre is as $n_{\text{halo}} (r_{\text{peri}})\sim n_{\text{gas}}\sigma^2/v^2_\text{peri}$, where $\sigma$ is the stellar velocity dispersion of the satellite, $v_{\text{peri}}$ is the relative motion of the galaxy at $r_{\text{peri}}$, and $n_{\text{gas}}$ is the average gas density in the inner region of the satellite. 

It was previously shown that this approach may underestimate the halo density by the factor of $\sim 5$ compared with the simulations of ram-pressure stripping \citep[][]{Gattoetal2013}. Hence, these data has not been included in the fit. 

In Figure~\ref{fig:rampressure}, the constraints from~\citet{Putmanetal2021} that predict $n_e\ga10^{-5}$~cm$^{-3}$ within the assumed virial radius are compared with the best-fitting profile obtained in this work. Wherever these constraints require a higher density than that modelled, it can be speculatively concluded that the corresponding satellite has undergone ram-pressure stripping in a region of above-average density, since the profile only represents the background density.

Thus, it can be evidently claimed that the model is consistent with the ram-pressure stripping data it was fitted to.

\begin{figure}
	\includegraphics[width=\columnwidth]{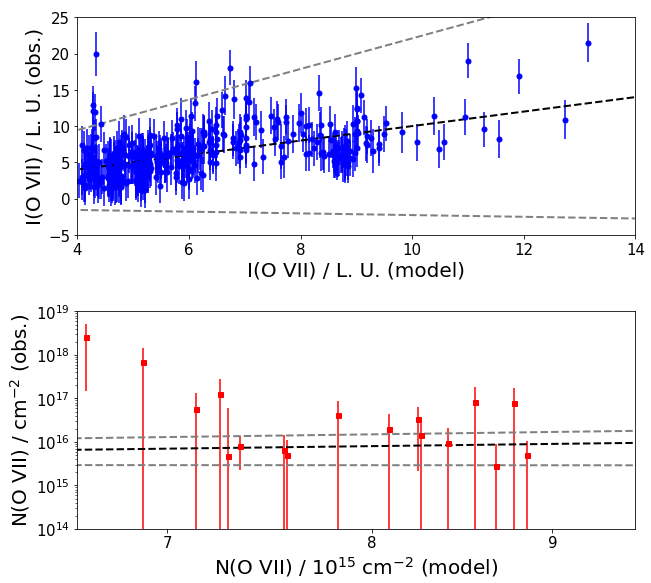}
    \caption{Observed \ion{O}{vii} line strengths (top) and column densities (bottom) compared with the modelled values. The black (gray) dashed lines represent the best-fitting value (68 per cent CL allowed region).}
    \label{fig:intensity_columndensity}
\end{figure}

\paragraph*{Oxygen spectra.} In Figure~\ref{fig:intensity_columndensity}, the modelled and observed \ion{O}{vii} line intensities and column densities are compared. The modelled values agree within $1\sigma$ with 429 (17) out of 431 (18) oxygen emission (absorption) observations. 

The modelled column densities of $\log(N_{\text{\ion{O}{vii}}}/\text{cm}^{-2})\simeq(15...16)$ agree well with that predicted in the recent works by~\citet{Voit2019} and \citet{Kaaretetal2020}, where, respectively, $\log(N_{\text{\ion{O}{vii}}}/\text{cm}^{-2})\sim16$ and $\log(N_{\text{\ion{O}{vii}}}/\text{cm}^{-2})\simeq(15.7...16.1)$ (for $|b|>30\degr$) is reported. 

Herewith, the modelled line strengths are characterized by the large degree of uncertainty. Remarkably, $\sim 85~(10)$ per cent of the lower (upper) limit is due to the LB density parameter uncertainty. 

Generally, it is however can be concluded that the model self-consistently describes oxygen spectra. 

\paragraph*{Dispersion measures.} The DM is defined as follows:
\begin{equation}
    \text{DM} = \int\limits_{0}^{D} n_e(s) ds,
\end{equation}
where $s$ is the sight line coordinate and $D$ is the distance to the pulsar.

The fitting procedure used the results from \citet{AndersonBregman2010}, where the value of $\simeq23$ cm$^{-3}$ pc \citep[inferred from the observation by][]{Manchester2006} was derived as an estimate for non-Galactic contribution to the DM of pulsars in the LMC direction. This is also consistent with the DMs of the LMC pulsars discovered later by \citet{Ridley2013}. 

Following the work of \citet{AndersonBregman2010}, the distance to the LMC of $D=50$ kpc is assumed. The corresponding DM is $\simeq 16.6\,(9.3...22.8)$ cm$^{-3}$ pc. This indicates model self-consistency, because the electron density profile is responsible for the background halo density, and therefore the modelled DM is expected not to be greater than that observed.

This result also accords those from \citet{Nugaevetal2015}, where DM $\simeq (19.8...22.9)$ cm$^{-3}$ pc is calculated assuming the halo density profile obtained in~\citet{Feldmann2013}, and from \citet{Yamasaki2020}, where DM $\simeq (17...21)$ cm$^{-3}$ pc is estimated assuming the model presented by~\citet{Yao2017}.

\subsubsection{Temperature assumption}\label{subsubsec:temperature}
It is reasonable to examine the temperature assumption in the context of the gas hydrostatic equilibrium. The corresponding equation is:
\begin{equation}
    \frac{dP}{dr} = -G \frac{M(r)\rho(r)}{r^2},
\end{equation}
where $P$ and $\rho$ are, respectively, the gas pressure and density, $G$ is the gravitational constant and $M(r)$ is the total mass within the galactocentric radius of $r$. Supposing that the gas is ideal, $P = \rho k_{B} T/\mu m_{p}$ and then \citep[see e. g.][]{Makinoetal1998}:
\begin{equation}
    \frac{d\ln{\rho}}{dr} = -G\frac{\mu m_{p}M(r)}{k_{B}T r^2},
    \label{eq:densitylogderiv}
\end{equation} 
where $\mu\simeq0.59$ is the mean mass per particle in the units of $m_{p}$ (mass of the proton), $k_{B}$ is the Boltzmann constant.
The present analysis assumes that the gas density follows the $\beta$-profile (see Equation~\ref{eq:beta-profile-simplified}), and therefore:
\begin{equation}
    T = \frac{\mu m_{p} GM(r) \left(1+(r/r_{c}\right)^2)}{3\beta k_{B}r \left(r/r_{c}\right)^2}
\end{equation}
At the radius of $r=R_{\text{vir}}\gg r_c$:
\begin{equation}
    T \simeq \frac{\mu m_{p} GM_{\text{vir}}}{3\beta k_{B} R_{\text{vir}}} = 1.59\times 10^{6}\left(\frac{\beta}{0.337}\right)^{-1}\left(\frac{M_{\text{vir}}}{1.17\times10^{12}~\text{M}_{\sun}}\right)^{2/3}\text{K}
    \label{eq:temp_hydrostatic}
\end{equation}
According to the recent MW mass measurements \citep[see][]{Wang2020},  $M_{\text{vir}}=(0.5...2.0)\times10^{12}~\text{M}_{\sun}$. In addition, in this self-consistency test, only the logarithmic derivative of the gas profile is constrained (see Equation~\ref{eq:densitylogderiv}), and thus the normalization $\alpha$ should be considered as a free parameter, which results in $\beta=(0.186...0.511)$ (see Figure~\ref{fig:betaparameters}).
Thus, the corresponding temperature is $T \simeq (0.6...3.0)\times10^{6}\text{ K}$, and the temperature assumption of $T = 2\times 10^{6}\text{ K}$ can be concluded to be quite reasonable.

\begin{figure}
	\includegraphics[width=\columnwidth]{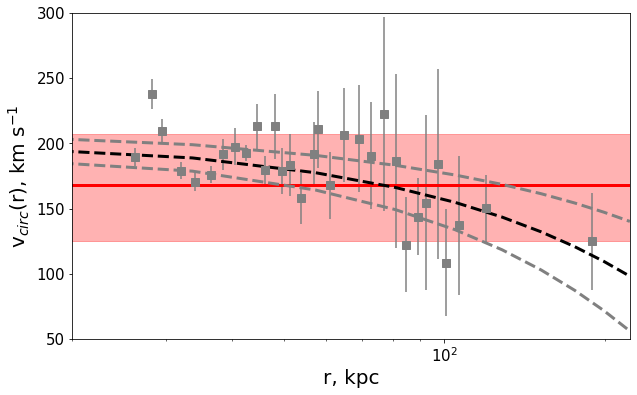}
    \caption{Observed circular velocities of non-disc objects from~\citet{Bhattacharjee2014} (gray points with error bars) compared with the modelled value. The black (gray) dashed lines represent the linear fitting (68 per cent CL allowed region), the red solid line (red-shaded area) represents the modelled circular velocity (68 per cent CL allowed region).}
    \label{fig:circular}
\end{figure}

From Equation~\ref{eq:temp_hydrostatic}, the circular rotation velocity of $\sqrt{GM(r)/r}$ implied by the model can be constrained at the galactocentric distances of $r\ga 20 \text { kpc}$, where the true gas distribution is nearly spherical and the modelled profile slope is not distorted by the additional contribution from the disc component not considered by the current analysis. Figure~\ref{fig:circular} compares the modelled values with the observed circular velocity for non-disc objects from~\citet{Bhattacharjee2014}. As can be seen, the model is consistent with these observations within the uncertainty, and this also validates the temperature assumption.

It also can be concluded that the electron density profile slope parameter $\beta$ is related to the gas temperature $T$ as follows \citep[see][]{Troitsky2017,BregmanAnderson2018}:
\begin{equation}
    \beta = T_{\text{rot}}/T = \frac{\mu m_{p} \sigma^2}{k_{B}T} \simeq 0.715 \left(\sigma/100\text{ km s}^{-1}\right)^{2} \left(T/10^6 \text{ K}\right)^{-1},
\end{equation}
where $T_{\text{rot}}$ is the thermal energy associated with the circular rotation velocity, and $\sigma$ is the velocity dispersion of galactic objects. The best-fitting slope parameter of $\beta = 0.337^{+0.174}_{-0.151}$ and the assumed temperature of $\log(T_{\text{halo}}/\text{K}) = 6.3$ infer the velocity dispersion of $\sigma_{\text{model}} \simeq (75...122) \text{ km s}^{-1}$, which agrees well with the observed values $\sigma_{\text{obs}} \simeq (90...120) \text{ km s}^{-1}$ in the range of galactocentric distances from $r\simeq 20$~kpc up to $r\simeq 80$~kpc \citep[][]{NestiSalucci2013}. 

At the galactocentric distances of $r\la20$~kpc, the axial symmetry dominates. Hence, to test the temperature assumption in this region, one should compare axial symmetric Galactic potential model predictions with that of the spherical model presented here.

\begin{figure}
	\includegraphics[width=\columnwidth]{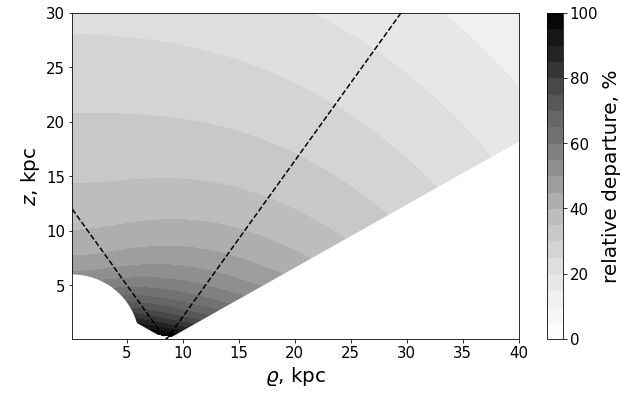}
    \caption{Gravitational force per unit mass relative departure (from the expected value) implied by the spherical model of the gas near hydrostatic equilibrium. Dashed lines correspond to the median absolute galactic latitude in the spectroscopic sample. The area beyond the model applicability region is excluded. See Subsections~\ref{subsubsec:limitations} and~\ref{subsubsec:temperature} for a detailed discussion.}
    \label{fig:force}
\end{figure}

Figure~\ref{fig:force} illustrates the relative departure of the gravitational force per unit mass predicted by the spherical model (assuming $T = 2\times 10^{6}\text{ K}$ and $\beta = 0.337$) from the axial symmetric model prediction. This relative departure~is:
\begin{equation}
    \text{rel. dep. } \mathcal{F} = \left( \left(\frac{\mathcal{F} \times \varrho/r - \mathcal{F}_{\varrho}}{\mathcal{F}_{\varrho}}\right)^2 + \left(\frac{\mathcal{F} \times z/r - \mathcal{F}_{z}}{\mathcal{F}_{z}}\right)^2 \right)^{1/2},
\end{equation}
where $\mathcal{F}$ is the gravitational force per unit mass of $GM(r)/r^2$ calculated within the spherical model using Equation~\ref{eq:densitylogderiv} (to be consistent with previous studies, $r_c = 3$~kpc is assumed), $\varrho$ and $z$ are cylindrical coordinates, and $\mathcal{F}_{\varrho}$ and $\mathcal{F}_{z}$ are the corresponding force components calculated within the axial symmetric \texttt{MWPotential2014} model using \texttt{galpy} package \citep[][]{JoBovy2015}. 

Since the force predicted by the spherical model is proportional to the assumed temperature and underestimates the actual value near the disc, the relative departure can be interpreted as the temperature relative underestimation. Not surprisingly, this underestimation increases near the MW disc. Within $z\la7$~kpc, the relative departure is $\simeq (50...100)$~per~cent, which is larger than $\beta$ uncertainty of $\simeq 50$~per~cent. 

The emission-weighted temperature for the best-fitting parameters (statistical weight is $\propto n_e^2 Z$, which reflects the contribution to the line intensity) along the sight line of the median absolute galactic latitude $|b|\simeq55\degr$ is thus $T_{\text{em}}\simeq3.1\times10^{6}$~K, which is $\simeq1.5$ times larger than that assumed, but is still consistent with the spherical model taking into account $\beta$ uncertainty.

However, it should be acknowledged that near the disc plane, there is a conflict between the spherical model of the gas near hydrostatic equilibrium and the temperature assumption. The possible explanation of this issue is non-thermal pressure support in this region, e.~g.~from cosmic rays and turbulent motion. According to Figure~\ref{fig:force}, the corresponding pressure gradient at $z\la7$~kpc should be of the order of the hot gas pressure gradient.

\subsubsection{Spherical symmetry assumption}\label{subsubsec:sphsymm}
To test the spherical symmetry assumption, the spectroscopic sample was divided into two sub-samples corresponding to the right ($l < 180\degr$) and left ($l > 180\degr$) Galactic hemispheres. According to the symmetry assumption, the fitting results are expected to be insensitive to the choice of the hemisphere. Thus, the model has been fitted to the both left and right sub-samples separately using the best-fitting parameter set (Table~\ref{tab:Results}) as an initial guess. 

\begin{figure}
	\includegraphics[width=\columnwidth]{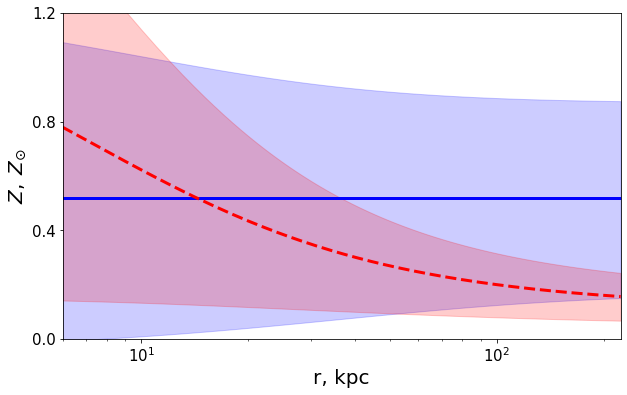}
    \caption{Metallicity profile of the circumgalactic gas with the corresponding $1\sigma$ allowed regions obtained in this work using observations from $l<180\degr$~(red, dashed) and $l>180\degr$~(blue, solid). See Sub-subsection~\ref{subsubsec:sphsymm}.}
    \label{fig:sphsymm}
\end{figure}

The results are presented in Table~\ref{tab:Symmetry}. As can be seen, the derived electron density and the LB density are statistically identical within the uncertainty. However, there is a difference $\ga1\sigma$ between the metallicity profile parameters and $\simeq 1\sigma$ between the corresponding profiles at large galactocentric distances (Figure~\ref{fig:sphsymm}). 

It is difficult to appropriately interpret this result, since it could have been affected by many factors, including the real asymmetry of physical properties of the halo and/or the LB, and completely methodical features of the sample, such as asymmetry in the number of observations and their accuracy.

Interestingly, the LB asymmetry finding (that however has not enough statistical confirmation provided by this work and must be considered with cautious) qualitatively accords the data obtained in~\citet{Lallement2003}, where 3D density maps of the local interstellar gas are presented. The authors report on the longer characteristic path in the dense gas clouds for $l>180\degr$~than that for $l<180\degr$, which in terms of spherical model of the LB not surprisingly converts into the higher best-fitting density.    

In general, it should be concluded that a potential bias associated with the halo spherical asymmetry cannot be neither rejected nor confirmed, but the electron density profile herewith can be concluded not to be crucially distorted by the asymmetry. 

\begin{table}
	\centering
	\caption{Results of the spherical symmetry test (see Sub-subsection~\ref{subsubsec:sphsymm}). The top and the bottom lines of the table represent, respectively, results the right ($l < 180\degr$) and left ($l > 180\degr$) regions of galactic coordinates.}
	\label{tab:Symmetry}
    \begin{tabular}{l c c c c c}
        \hline 
        \hline
        $\alpha$, cm$^{-3}$ kpc$^{3\beta}$ & $\beta$ & $a$, $Z_{\sun}$ & $d$, kpc & $n_{\text{LB}}$, cm$^{-3}$\\
        \hline
        7.2$^{+1.7}_{-2.1}\times10^{-3}$ & 0.34$^{+0.04}_{-0.03}$ & 0.11$^{+0.06}_{-0.06}$ & 25$^{+7}_{-12}$ & 0$^{+9}_{-9}\times10^{-3}$ \\
        \hline
        5.5$^{+1.5}_{-1.9}\times10^{-3}$ & 0.32$^{+0.04}_{-0.03}$ & 0.52$^{+0.35}_{-0.35}$ & 0$^{+8}_{-19}$ & 5$^{+5}_{-15}\times10^{-3}$ \\
        \hline
        \end{tabular}
\end{table}

\subsection{Implications for the MW}\label{subsec:implications}
In this subsection, estimates of the MW physical properties inferred by the results of this work are discussed.
\subsubsection{Missing baryons}
It is interesting to explore the electron density profile obtained here in the context of the missing baryon problem. The total hot gas mass can be expressed as follows:
\begin{equation}
\begin{gathered}
M_{\text{gas}} = \int\limits_{R_{\text{in}}}^{R_{\text{vir}}} \mu m_{p} n_{\text{halo}} (r) 4\pi r^2 dr \Rightarrow\\ \frac{M_{\text{gas}}}{10^{5}~\text{M}_{\sun}} \simeq  1.2 \left(\frac{\alpha/(1-\beta)}{10^{-3} \text{ cm$^{-3}$ kpc$^{3\beta}$}}\right)\left(\frac{r}{1\text{ kpc}}\right)^{3(1-\beta)}\bigg|^{r = 223\text{ kpc}}_{r = 6\text{ kpc}},
\end{gathered}
\end{equation}
where $n_{\text{halo}} \simeq 1.9 n_e$ is the total number density of the gas. The derived mass is $M_{\text{gas}} \simeq 5.5^{+3.2}_{-3.1} \times 10^{10}~\text{M}_{\sun}$, which is consistent within 1$\sigma$ with the constraints reported by \citet{AndersonBregman2010,MillerBregman2013,MillerBregman2015,Nuza2014,Kaaretetal2020}. 

To estimate the missing baryon mass for the MW, the approach presented by~\citet{AndersonBregman2010} is followed. The total MW baryon mass is divided into the observed and missing components:
\begin{equation}\label{eq:Mbtotal}
    M_{\text{obs}} + M_{\text{miss}} = M_{\text{b}} = f_b M_{\text{vir}},
\end{equation}
where $f_b=\Omega_b/\Omega_m\simeq0.15$ is the cosmic baryon fraction \citep[][]{Planck2015}. For the galaxies in the near-MW mass range there is a correlation \citep[see][]{AndersonBregman2010} between the observed baryon mass and the total mass, that can be parametrized as follows:
\begin{equation}\label{eq:Mbobs}
    M_{\text{obs}}/M_{\text{vir}} \simeq 0.04 \left(M_{\text{vir}}/2\times10^{12}~\text{M}_{\sun}\right)^{1/2}
\end{equation}
Combining Equation~(\ref{eq:Mbtotal}) and~(\ref{eq:Mbobs}), the missing mass can be estimated as:
\begin{equation}
     M_{\text{miss}} \simeq M_{\text{vir}} \left(0.15 -  0.04 \left(M_{\text{vir}}/2\times10^{12}~\text{M}_{\sun}\right)^{1/2}\right)
\end{equation}
Since the virial mass assumed in this work is $M_{\text{vir}}=1.17\times10^{12}~\text{M}_{\sun}$, the missing mass is $M_{\text{miss}} \simeq 14.0 \times 10^{10}~\text{M}_{\sun}$. Hence, the hot halo gas relative contribution can be estimated as $\simeq 39\,(17...61)$ per cent, which implies that the hot gas may hold a significant amount of the missing matter, comparable with the observed baryon mass in the MW and even larger than it was previously thought, see e.~g. \citet{AndersonBregman2010}, where $\simeq(6...13)$ per cent is reported, and \citet{LiBregman2018}, where hot baryons were concluded to be insufficient to explain the missing baryons in massive spiral galaxies (their contribution was estimated as only $\sim20$ per cent of the missing mass).

However, it is important to keep in mind that this approach assumes the MW to obey the correlation law~(\ref{eq:Mbobs}) and does not consider any direct observations to estimate the observed MW baryon mass, which could significantly distort the results. This makes the resulting estimate of the fraction of missing baryons held in CGM rather qualitative than quantitative. 

Herewith, the recent estimates of the total mass of the Galactic bulge and disc \citep[][]{Korol2019} and cool ($\log(T/\text{K})\sim4$) circumgalactic gas \citep[][]{Stern2016} imply, respectively, $M_{\text{disc+buldge}}\simeq7.8^{+1.4}_{-1.8}\times 10^{10}~\text{M}_{\sun}$ and $M_{\text{cool}}\simeq1.3^{+0.4}_{-0.4}\times 10^{10}~\text{M}_{\sun}$. Together with the the current analysis, the total MW baryon mass can be estimated as $M_{\text{obs}}\simeq14.6^{+3.7}_{-3.6}\times 10^{10}~\text{M}_{\sun}$. This agrees with the assumed $f_b M_{\text{vir}}\simeq17.6\times 10^{10}~\text{M}_{\sun}$ within $1\sigma$ and thus indicates that CGM could contain all the missing baryons of the Galaxy.

\subsubsection{X-ray luminosity \& accretion rate}
Following the works of \citet{MillerBregman2015,Troitsky2017}, here, the MW X-ray luminosity inferred by the best-fitting results is examined. The following expression for the cooling time \citep[see][]{FukugitaPeebles2006}, which determines whether or not the hot gas halo is stable at present-day stage in the MW evolution, is used: 
\begin{equation}
    t_{\text{cool}} = \frac{1.5 n_{\text{halo}} k_{\text{B}} T}{\Lambda (T, Z) n_e (n_{\text{halo}} - n_e)} \simeq 2.1 \times \frac{1.5 k_{\text{B}} T}{\Lambda (T, Z) n_e }, 
\end{equation}
where $\Lambda (T, Z)$ is the bolometric cooling rate \citep[see][Figure 13]{SutherlandDopita1993}. In this study, a constant temperature of $\log(T_{\text{halo}}/\text{K})=6.3$ is assumed, but metallicity is not a constant. Thus, the results obtained by~\citet{SutherlandDopita1993} are used to derive a linear parameterization of $\Lambda (T_{\text{halo}},Z)$ in the region of $\text{[Fe/H]}\simeq\log(Z/Z_{\sun})\simeq(-0.5 ... 0)$, corresponding to the best-fitting metallicity profile values (it should be kept in mind, however, that in the original work, a constant chemical composition was assumed):
\begin{equation}\label{eq:Lambda}
    \log\left(\Lambda/1 \text{ erg cm$^3$ s$^{-1}$}\right) \simeq -22.2 + 0.6\log\left(Z/Z_{\sun}\right)
\end{equation}
The inferred cooling time is:
\begin{equation}\label{eq:tcool}
    t_{\text{cool}}(r) \simeq 4.4\times\left(Z(r)/Z_{\sun}\right)^{-0.6} \left(n_e(r)/10^{-4}\text{ cm}^{3}\right)^{-1}\text{ Gyr}
\end{equation}
From the condition $t_{\text{cool}} = 13.8$ Gyr, a corresponding cooling radius of $R_{\text{cool}}\simeq105\,(34...171)$~kpc is derived. From the cooling time radial function~(\ref{eq:tcool}), the current MW accretion rate can be estimated as follows \citep[see][Equation~18]{MillerBregman2015}:
\begin{equation}
    \Dot{M} = \int\limits^{R_{\text{cool}}}_{R_{\text{in}}} \frac{\mu m_{p} n_{\text{halo}}(r)}{t_{\text{cool}}(r)} 4\pi r^2 dr
\end{equation}
The corresponding (0.5...2.0) keV band luminosity of the MW:
\begin{equation}
    L_{X} = 0.412 \times \Dot{M} \frac{1.5k_{\text{B}}T}{\mu m_{p}},
\end{equation}
where 0.412 is a conversion coefficient. The resulting luminosity is $L_{X}\simeq 2.0^{+4.6}_{-2.6}\times 10^{40}\text{ erg s}^{-1}$. As can be seen, the luminosity is very poorly constrained in this work, partly because of the method of its calculation (it must be once again emphasized that a linear parameterization~(\ref{eq:Lambda}) was used for the variable metallicity, while in the original work a constant chemical composition was assumed). However, the upper luminosity limit of $L_{X} < 6.6\times10^{40} \text{ erg s}^{-1}$ is obtained, which is consistent with the total (0.5...2.0) keV band luminosity of $L_{X} \simeq (2...3) \times 10^{39}\text{ erg s}^{-1}$ modelled from the \textit{ROSAT} all-sky survey observations by \citet{SnowdenEgger1997,Wang1998}. 
It is also notable that the corresponding upper limit for the gas accretion rate $\Dot{M}<6.0~\text{M}_{\sun} \text{ yr}^{-1}$ is consistent with the constraints by \citet{Nuza2014}, where $\Dot{M} \simeq (6...8)~\text{M}_{\sun} \text{ yr}^{-1}$ is derived for all material within $R_{\text{vir}}$, and with estimates of the current Galactic star formation rate SFR $\simeq(0.7...5)~\text{M}_{\sun} \text{ yr}^{-1}$ \citep[see][]{Smith1978,Diehletal2006,Misitiotisetal2006,MurrayRahman2010,RobitailleWhitney2010,ChomiukPovich2011}.

\subsection{Final remarks}\label{subsec:finalremarks}
The shape of the metallicity profile obtained here should be considered cautiously, since the argumentation to derive it was completely qualitative. This work has not considered any other possible parameterizations for the profile and thus it cannot be claimed that the proposed one is optimal. 

Moreover, the proposed parameterization for the metallicity implies a noticeably more strict constraints at large radii (beyond 50~kpc). Physically, it seems to be a confusing result, since the observational data is obviously dominated by the dense central region of the halo \citep[within 50 kpc, see e. g.][]{BregmanAnderson2018}. This is likely to have been caused by the metallicity profile shape, since in the considered parameterization, the variations of the parameters imply larger metallicity variations at smaller radii. The metallicity constraint should be thus interpreted with caution, since the analysis is almost insensitive to the metallicity values at large galactocentric distances due to the fact that this region contributes insignificantly to the spectroscopic measurements, and this constraint is therefore substantially model-dependent. This also can be interpreted as an indication that a more optimal profile shape is needed.

None the less, this study, according to the author's knowledge, for the first time, has introduced a physical parameterization of the metallicity radial dependence and constrained it from the observational data. Further research is suggested to expand the range of possible parameterizations in order to choose the optimal one and to derive more precise constraints on the metallicity, in particular, to explore whether or not it is reasonable to take into account the metallicity gradient in terms of one or another calculation.

It should be also kept in mind that the fitting procedure assumed the contribution from spectroscopic and ram-pressure stripping observations to be statistically equal despite the noticeable difference in the number of measurements, 7 compared to 449. In addition, ram-pressure stripping was assumed the only significant mechanism of dwarf satellites' gas loss, which was motivated by qualitative estimates but was not conclusively verified. Also, this work did not consider an additional systematic uncertainty in the analysed ram-pressure stripping data sample associated with the orbital parameters of the dwarf galaxies. Modifying of this approach might lead to the result different from that obtained here, but this is beyond the scope of the current analysis.

\section{Conclusions}\label{sec:conclusions}
In this work, a joint analysis of the ram-pressure stripping and \ion{O}{vii} emission and absorption associated with the Milky Way's circumgalactic medium has been presented. The sample was combined from the data obtained in the works by \citet{GrcevichPutman2009,Gattoetal2013,Salemetal2015} for ram-pressure stripping, \citet{MillerBregman2013,Fang2015} for oxygen absorption and \citet{HenleyShelton2012} for oxygen emission. In order to focus on the data due to the Milky Way's hot gaseous halo, a filtering procedure was applied to the spectroscopic sample. The sight lines of high \ion{H}{i} column densities \citep[][]{Westmeier2018}, as well as the regions near the galactic plane ($|b|<30\degr$) and near the Fermi Bubbles were excluded from the consideration. This filtering reduced the sample to 431 (18) observations of \ion{O}{vii} emission (absorption) and 7 electron density constraints from ram-pressure stripping studies.

Then, a parametric model of the Milky Way's circumgalactic medium has been developed by modifying the model presented in~\citet{MillerBregman2013,MillerBregman2015}. A spherical isothermal halo was considered, and the isotropic background from the Local Bubble and any other non-halo contribution was subtracted. To be consistent with previous studies, for the electron density, a spherical $\beta$-profile reduced to the power law at galactocentric distances of interest was assumed. 

The advantage of the analysis presented here over previous efforts on the oxygen spectra modelling is that, for the first time, a physically motivated parametric metallicity profile was introduced. To derive this profile, the gas was considered to be divided into the primordial (of zero metallicity) and late (of constant near-solar metallicity) components, and the distribution of the latter component in the Galactic gravitational potential primarily determined by the dark matter density profile was qualitatively described. This allowed to constrain the gas chemical composition gradient as well as to resolve a discrepancy between the results of modelling of spectroscopic and ram-pressure stripping data. However, to claim that the discrepancy is conclusively resolved, in further studies, it is required to re-simulate the ram-pressure stripping process using the improved orbital parameter values.

Then, using the previously derived constraints on the gas electron density \citep[from the analysis of ram-pressure stripping and the LMC pulsars dispersion measures, see][respectively]{BlitzRobishaw2000, AndersonBregman2010}, its metallicity and the LB density \citep[from oxygen emission observations, see][]{MillerBregman2015}, a step-by-step fitting procedure was developed in order to easily explore the 5-dimensional parameter space (the $\beta$-profile slope and normalization parameter, the metallicity profile slope and asymptotic parameter, and the LB density parameter). With the aim of taking into account methodological inaccuracy of the emission lines measurements and finding an acceptable $\chi^2$, the systematic uncertainty of $\simeq40$ per cent of the median line intensity was added in quadrature to these data. 

An acceptable reduced $\chi^2$ value of $\chi^2/\text{dof}\text{ (dof)}\simeq1.43\,(458)$ has been found, misleadingly indicating an under-fit to the ram-pressure stripping observations. The latter has been concluded to be due to the uncertainties symmetrization, which has a sufficient affect on the contribution of these data. The actual minimized value of $\chi^2_{\text{true}}/\text{dof}\text{ (dof)}\simeq1.02\,(458)$ was argued for (Subsection~\ref{subsec:selfconsistency}).

The electron density profile is found to be rather flat with the slope parameter of $\beta\simeq(0.31...0.38)$, corresponding to the behaviour of $\propto r^{-(0.9...1.1)}$ at large radii ($r\gg1$ kpc). This profile is flatter than that reported in the works by~\citet{MillerBregman2013,MillerBregman2015,LiBregman2018,Kaaretetal2020} (based on the spectroscopic data exclusively), which is likely to be due to the taking into account a metallicity gradient but apparently is not due to the ram-pressure stripping data contribution. However, the profile accords with that obtained in \citet{MillerBregman2015} within $1\sigma$. Also, the obtained profile slope agrees with that of other nearby massive spiral galaxies within $(1...2)\sigma$ \citep[][]{Bogdan2013,LiBregman2018}. The profile is however considerably flatter than that typical for observed external galaxies for which metallicity can be measured.

The metallicity profile and the LB density parameter are poorly constrained by this work. Due to the large uncertainties, it cannot be conclusively claimed whether or not the metallicity gradient is significant. The metallicity in the outer part of the halo ($r\ga50$ kpc) has been constrained as $Z\simeq(0.1...0.7)~Z_{\sun}$ assuming the introduced profile shape and a constant ionization fraction of $f=0.5$ (it should be emphasized that physically, the analysis is almost insensitive to the metallicity values at large galactocentric distances, and this constraint is model-dependent, see Subsection~\ref{subsec:finalremarks}). None the less, this approach relaxes the constant-metallicity assumption and thus allows to self-consistently describe the spectroscopic data together with ram-pressure observations. It also provides direct constraints on the metallicity, which might be a useful tool for future research. 

It is also notable that despite the fact that LB density parameter constraints (observational and derived from the model) are very inaccurate, the fitting procedure was shown to be stable with respect to $\leq25$ per cent initial variations of this parameter, and the best-fitting density profile is practically insensitive even to the neglecting of the LB contribution. These can be considered as additional arguments for the results reliability.

All the constraints derived here should be emphasized to characterize only the background density of the gas with no regards to the halo substructure and to be inapplicable at the galactocentric distances beyond $6\text{ kpc}\leq r \leq 223\text{ kpc}$ and the galactic latitudes and longitudes, respectively, beyond $|b|>30\degr$ and $10\degr<l<350\degr$, since they were obtained using the data collected (or, speaking about the outer cutoff radius, assumed to be collected) only from this region. The gas distribution obtained here describes the gas primarily at large galactocentric distances and does not reflect the actual gas distribution in the near-disc region ($r\la 20 \text{ kpc}$), where axial symmetric disc-associated gas distribution dominates over the halo-associated spherical profile constrained here. Cautiously, or qualitatively, the results still can be used to describe the hot gas beyond the discussed region, but only with regard to the possibility of significant quantitative distortions. 

The model self-consistency and implications are discussed in detail in Section~\ref{sec:discussion}. Here, the results are briefly summed up, once again emphasizing that all the constraints presented below should be interpreted as rather qualitative than quantitative arguments for the model appropriateness and self-consistency.
\begin{enumerate}
    \item The test of spherical symmetry that was conducted by the fitting of the model to the observations from the left and right hemispheres of Galactic longitudes revealed a $\sim1\sigma$ difference in the metallicity profiles, and thus the symmetry should be assumed cautiously in further studies. This work does not provide a conclusive interpretation of this effect.
    \item The result agrees well with the lower limit for the volume-averaged halo density derived in the work by~\citet{BlitzRobishaw2000} and with the ram-pressure stripping constraints presented by~\citet{Putmanetal2021}, as well as with the estimates of the halo dispersion measure in the direction of the LMC obtained by~\citet{AndersonBregman2010,Nugaevetal2015,Yamasaki2020}.
    \item The temperature assumption is shown to be self-consistent in the context of the gas hydrostatic equilibrium in the MW halo. The temperature inferred from the results is $T \simeq 1.59\,(0.6...3.0)\times10^{6}\text{ K}$, while the assumed temperature is $T=2\times10^{6}\text{ K}$. Also, the modelled circular rotational velocity for non-disc objects is consistent with the observed values~\citep[][]{Bhattacharjee2014} within the uncertainty, and the velocity dispersion of the near-Galactic objects of $\sigma_{\text{model}} \simeq (75...122) \text{ km s}^{-1}$ inferred from the halo temperature assumption and the derived electron density profile slope parameter agrees well with that of $\sigma_{\text{obs}} \simeq (90...120) \text{ km s}^{-1}$ inferred from observations \citep[][]{NestiSalucci2013}. However, near the disc plane ($z\la7$~kpc), there is a conflict between the temperature and the hydrostatic equilibrium assumptions indicating that the non-thermal pressure support (e.~g. from cosmic rays or turbulent motion) could take place, with the pressure gradient of the same order as that of the hot gas.
    \item The total Milky Ways' hot gaseous halo mass implied by the results is $M_{\text{gas}}\simeq5.5^{+3.2}_{-3.1}~\text{M}_{\sun}$, which is consistent with the previous studies \citep[][]{AndersonBregman2010,MillerBregman2013,MillerBregman2015,Nuza2014,Kaaretetal2020} within $1\sigma$. It is argued however that the halo can be responsible for $\sim(17...100)$ per cent of the MW's missing baryon mass, which is larger than it was previously thought.
    \item The upper limit for the MW's (0.5...2) keV band luminosity $L_{X}<6.6\times10^{40}\text{ erg s}^{-1}$ as well as the corresponding mass accretion rate $\Dot{M}<6.0~\text{M}_{\sun}\text{ yr}^{-1}$ is in good qualitative agreement with that estimated from observations~\citep[][]{SnowdenEgger1997,Wang1998,Nuza2014}.
\end{enumerate}

Even with all the limitations and imperfections discussed above, this research has generally succeeded in providing a joint self-consistent analysis of ram-pressure stripping occurred in the Galactic halo and spectra of circumgalactic oxygen, making another step towards the understanding of physical properties of the circumgalactic medium.

Further work is required to improve the model presented here, specifically, by considering other possible physical parameterizations of the metallicity profile, taking into account the halo substructure features and asymmetry and optical depth corrections and, as a result, expanding the model applicability region and the data sample under consideration.

\section*{Acknowledgements}
The author is indebted to S. V. Troitsky for the conceiving of the original idea and for many tremendously helpful comments and discussions on this work and to P. Mikushin and A. Trifonov for interesting discussions and helpful remarks on data processing. 

The author also thanks the anonymous referee for valuable comments and suggestions.

This study made use of the \texttt{Python} packages \texttt{matplotlib} \citep[][]{matplotlib}, \texttt{numpy} \citep[][]{numpy}, \texttt{galpy} \citep[][]{JoBovy2015} and \texttt{scipy} \citep[][]{scipy}. 

This work is supported in the framework of the State project “Science” by the Ministry of Science and Higher Education of the Russian Federation under the contract 075-15-2020-778.

\section*{Data Availability}

The data underlying this study are available in the articles referred to in the text. The reduced data sample which was the model fitted to and the original code will be shared by the author upon reasonable request.



\bibliographystyle{mnras}
\bibliography{example} 








\bsp	
\label{lastpage}
\end{document}